\documentclass[10pt,journal,compsoc]{IEEEtran}
\usepackage{cite}
\usepackage{graphicx}
\usepackage{balance}  % for  \balance command ON LAST PAGE  (only there!)
\usepackage{color}
\usepackage{amsfonts,amssymb,amsmath,bm}
\usepackage{makecell}
\usepackage{subfigure}
\usepackage{multirow}
\usepackage[ruled,vlined]{algorithm2e}
\usepackage{booktabs}
\usepackage{hyperref}
\usepackage{threeparttable}

\newtheorem{example}{Example}

\begin{document}
\title{CollaborER: A Self-supervised Entity Resolution Framework Using Multi-features Collaboration}

\author{Congcong~Ge,
        Pengfei~Wang,
        Lu~Chen,
        Xiaoze~Liu,\\
        Baihua~Zheng,
        Yunjun~Gao,~\IEEEmembership{Member,~IEEE}

\IEEEcompsocitemizethanks{
        \IEEEcompsocthanksitem C. Ge, L. Chen, X. Liu, and Y. Gao (Corresponding Author) are with the College of Computer Science, Zhejiang University, Hangzhou 310027, China, E-mail:\{gcc, luchen, xiaoze, gaoyj\}@zju.edu.cn.
        \IEEEcompsocthanksitem P. Wang is with the School of Software, Zhejiang University, Hangzhou, China, E-mail: wangpf@zju.edu.cn.
        \IEEEcompsocthanksitem B. Zheng is with the School of Computing and Information Systems, Singapore Management University, Singapore 178902, Singapore, E-mail: bhzheng@smu.edu.sg.
}

% \thanks{Manuscript received April 19, 2005; revised August 26, 2015.}
}

\IEEEtitleabstractindextext{
\begin{abstract}
Entity Resolution (ER) aims to identify whether two tuples refer to the same real-world entity and is well-known to be labor-intensive.
It is a prerequisite to anomaly detection, as comparing the attribute values of two matched tuples from two different datasets provides one effective way to detect anomalies.
Existing ER approaches, due to insufficient feature discovery or error-prone inherent characteristics, are not able to achieve a stable performance. 
In this paper, we present \textsf{CollaborER}, a self-supervised entity resolution framework via multi-features collaboration.
It is capable of (i) obtaining reliable ER results with zero human annotations and (ii) discovering adequate tuples' features in a fault-tolerant manner.
\textsf{CollaborER} consists of two phases, i.e., automatic label generation (ALG) and collaborative ER training (CERT).
In the first phase, ALG is proposed to generate a set of positive tuple pairs and a set of negative tuple pairs. ALG guarantees the high quality of the generated tuples, and hence ensures the training quality of the subsequent CERT.
In the second phase, CERT is introduced to learn the matching signals by discovering graph features and sentence features of tuples collaboratively.
Extensive experimental results over eight real-world ER benchmarks show that \textsf{CollaborER} outperforms all the existing unsupervised ER approaches and is comparable or even superior to the state-of-the-art supervised ER methods.
\end{abstract}

}

\maketitle

\IEEEdisplaynontitleabstractindextext

\IEEEpeerreviewmaketitle

\IEEEraisesectionheading{\section{Introduction}\label{sec:introduction}}

\IEEEPARstart{D}{ue} to widespread data quality issues~\cite{dataQuality2012}, anomaly detection has received tremendous attention in diverse domains.
It aims to find anomalous data in a dataset.
Many studies~\cite{ZongSMCLCC18, AudibertMGMZ20} focus on anomaly detection based on the information provided by a single dataset.
Differently, we would like to highlight that the anomaly detection problem can be facilitated by considering the information captured by different sources, since it is common that different data sources can provide information about the same real-world \emph{entity}~\cite{ERsurvey2012}.
Given two tuples from different relational datasets that refer to the same entity in real life, if they contain the same attribute but have contradictory values in the cell of the attribute, at least one of the values is an anomaly.

\begin{figure}[t]
\centering
\includegraphics[width=3.6in]{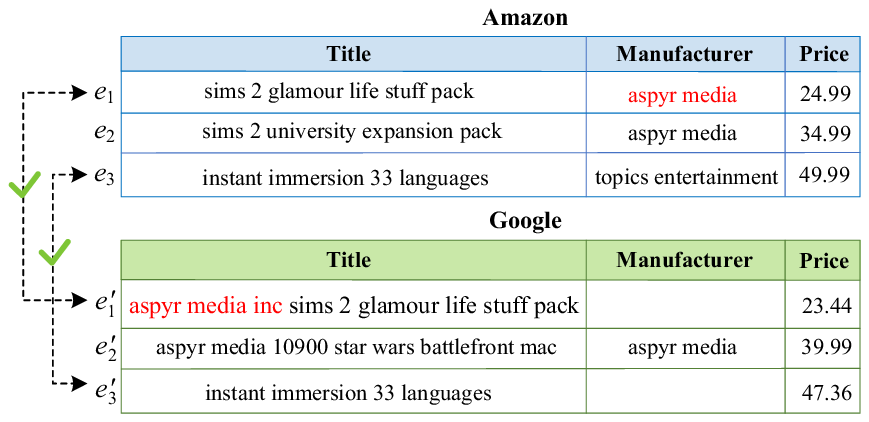}
\vspace{-3.5mm}
\caption{Example of using ER for anomaly detection}
\label{fig:example}
% \vspace{-2mm}
\end{figure}

\begin{example}\label{example:intro}
Figure~\ref{fig:example} depicts two sampled tables, each of which contains three tuples about products gathered from Amazon and Google, respectively.
% An ER strategy can be performed to detect anomalies in equivalent tuples.
In this figure, we assume the matched tuples (connected by tick-marked lines) that refer to the same real-world entity have been perfectly identified.
Since $e_{1}'$ and $e_{1}$ are matched tuples, it is expected that the two tuples share the same values for a given attribute.
However, after comparing the attribute values of $e_{1}$ with the attribute values of $e_{1}'$, we can easily spot an anomalous value w.r.t. the attribute \emph{Title} of tuple $e_{1}'$, which might be caused by data extraction errors.
Specifically, the value ``aspyr media inc'' of $e_{1}'$, corresponding to the attribute \emph{Manufacturer}, is wrongly extracted into the cell of attribute \emph{Title}.
%Since $e_{1}'$ and $e_{1}$ are matched tuples, it is easy to detect anomalies according to their contradictory values.

% It is obvious that both matched tuple pairs contain anomalies due to the contradictory values.
% In the pair $(e_1, e_{1}')$, the values in the cell of the attribute \emph{manufacturer} of tuples $e_1$ and $e_2$ can be regarded as anomalies.
% This is because, the attribute \emph{manufacturer} of tuple $e_1$ has value ``aspyr media'', while that of tuple $e_2$ has a ``NULL'' value.
%\baihua{Is this considered as an anomaly? I thought this is an example of incomplete data.}
% which is the same as value ``aspyr media inc'' of tuple $e_{1}'$, though stored under attribute \emph{title} for $e_{1}'$.
%For tuple $e_{1}'$, ``aspyr media inc'' is an anomaly data due to data drift, since it should locate in the cell of attribute ``manufacturer'' rather than appear in that of attribute ``title''.
%\baihua{Congcong, in this example, we assume the ER pairs have been identified already by some algorithms/methods, and we are trying to find anomaly/inconsistency by comparing the data of two entities of each pair? No need to include the link between $e_2$ and $e'_2$ in this example, as it is not required in this example and cause confusion (as I am thinking why no cross links between other pairs like $e_2$ and $e'_1$.}
\end{example}

Considering that it is impractical to assume all the matched tuples are known beforehand, in this paper, we focus on Entity Resolution (ER)~\cite{ERsurvey2012}, which aims to identify whether a tuple from one relational dataset and another tuple from a different relational dataset refer to the same real-world entity.
It is worth noting that reliable ER results are a prerequisite to ensure the quality of multi-source-based anomaly detection. This is because anomaly detection on top of false-aligned tuples is meaningless.
%Otherwise, anomalies might not be correctly detected due to the false-positive matches.

% To this end, anomalies can be detected by discovering inconsistent data from different data sources that refer to the same entity, as described in Example~\ref{example:intro}.
%
%\baihua{Shall we say that we can quantify/evaluate the quality/reliability of data by considering the information captured by different sources? Given a mapped pair of tuples that refer to one entity in real life, if a property of the entity appears in both tuples, it is more likely that the property is a true property; if we find something from one tuple contradicts something from the other tuple, the tuples might be anomalies?  }

Early ER approaches require rules~\cite{FanJLM09, MEMPQS017} or crowdsourcing~\cite{GokhaleDDNRSZ14, MarcusWKMM11, crowdER12}, which are impractical for matching real-world entities with literal or symbolic heterogeneity.
% Recently, machine learning (ML) has become an increasingly powerful tool for representing data with semantic vectors or capturing statistical features of data
Recently, embedding has become an increasingly powerful tool to encode heterogeneous entities into a unified semantic vector space, giving birth to various embedding-based ER techniques.
Current embedding-based solutions to ER mostly rely on either \emph{sentence features} or \emph{graph features}.
The former~\cite{AutoEM19,Ditto20,DeepMatcher18, DeepER18, ERIJCAI19, ActiveTransferER19,bertER2021} treats each tuple as a sentence and learns the tuple's embedding according to the contextual information contained in the sentence.
The latter~\cite{CreatingEmDI20,GraphER20} first constructs graphs to represent tuples and then learns matching signals of tuples based on the graph structure.
Despite the considerable ER performance on several benchmarks, identifying tuples referring to the same real-world entity, however, is still a challenging endeavor. The challenges are mainly two-fold, as listed below.

\vspace{0.05in}
\noindent
\textbf{Challenge \uppercase\expandafter{\romannumeral 1}:} \emph{Labor-intensive annotations for generating pre-matched tuples}.
% Although there exist some non-embedding-based ER approaches with the purpose of finding matching tuples in an unsupervised way, the ER performance is far from satisfactory~\cite{ZeroER20, MongeE96, Cohen00, BilenkoMCRF03}.
Embedding-based ER can achieve considerable results but typically requires a large number of labeled tuple pairs.
The annotating process is labor-intensive and hence restricts the scope of its applications in real-world ER scenarios.
% A straightforward way is to perform ER by an unsupervised approach without any annotations.
% Several ER approaches~\cite{ZeroER20, MongeE96, Cohen00, BilenkoMCRF03, zhang2020unsupervised, CreatingEmDI20} have tried to perform ER in an unsupervised way.
Although several ER approaches~\cite{ZeroER20, zhang2020unsupervised, CreatingEmDI20} have tried to perform ER in an unsupervised way without any annotation, their ER performance is far from satisfactory due to the error-sensitive nature.
A large body of research~\cite{adversarialAttack19, attackIJCAI18, GoodfellowSS14} has indicated that unsupervised methods can be easily misled/fooled or attacked since they do not include any supervision signal. Therefore, even slight erroneous data may lead to wrong results.
For example, ZeroER~\cite{ZeroER20}, the state-of-the-art unsupervised ER method, achieves poor performance on dirty datasets, as confirmed in the experiments to be presented in Section~\ref{exp:overall}.
Since real-world ER datasets usually incorporate various errors, it is challenging to apply unsupervised methods to solve real-world ER tasks directly.
% , as demonstrated by Example~\ref{example:unsupervised}.

% \begin{example}\label{example:unsupervised}
% Take ZeroER~\cite{ZeroER20}, the state-of-the-art unsupervised ER method, as an example.
% DBLP-ACM-clean and DBLP-ACM-dirty are two versions of DBLP-ACM, a widely-used real-world ER benchmark.
% Here, DBLP-ACM-clean does not contain any error; while DBLP-ACM-dirty contains various errors.
% %Various errors exist in DBLP-ACM (dirty).
% The experimental results show that ZeroER can achieve high accuracy on DBLP-ACM-clean.
% On the contrary, the accuracy drops dramatically (from 96\% to 63\%) when performing ER on DBLP-ACM-dirty, as verified in the experiments to be presented in Section~\ref{exp:overall}.
% \end{example}

\vspace{0.05in}
\noindent
\textbf{Challenge \uppercase\expandafter{\romannumeral 2}:} \emph{Insufficient feature discovery of the tuples for ER}.
Based on our preliminary study, neither sentence-based nor graph-based approaches is able to discover sufficient features of tuples to achieve high-quality ER results. To ease the understanding of these two types of approaches, we detail their respective strengths and limitations in the following.

%We first focus on illustrating the strengths and drawbacks of the sentence-based methods, where the embedding of a tuple is highly relevant to its serialized attribute values. The benefit of sentence-based methods is the fault tolerance for erroneous values caused by data drift when representing the semantic embeddings of tuples.
For the sentence-based methods, the embedding of a tuple is highly relevant to its serialized attribute values.
It is resilient to anomalous values caused by data extraction errors.
Take the tuple $e_{1}'$ in Figure~\ref{fig:example} as an example.
The attribute value of \emph{manufacturer} (i.e., ``aspyr media inc'') appears in a different place (as a part of attribute \emph{title} instead of \emph{manufacturer}), due to data extraction errors.
%, as described in Example~\ref{example:intro}.
The sentence-based methods treat the tuple $e_{1}'$ as a sentence \emph{``aspyr, media, inc, sims, 2, glamour, life, stuff, pack, 23.44''}.
In other words, ``aspyr media inc'' is still a part of the context of $e_{1}'$ and can provide effective information to learn the embedding of $e_{1}'$.
Despite the benefit, two main limitations exist in the sentence-based methods.
First, recent work~\cite{CreatingEmDI20} clarifies that, tuples are not sentences, and hence, treating a tuple blindly as a sentence loses a large amount of contextual information present in the tuple.
Second, they dismiss the rich set of semantics inherent among different tuples~\cite{CreatingEmDI20}.
To be more specific, they assume that different tuples are mutually independent.
On the contrary, it is common that different tuples share the same attribute values, and some common attribute values might appear in many tuples.
As shown in Figure~\ref{fig:example}, the attribute value ``aspyr media'' exists in both tuple $e_1$ and tuple $e_2$.

On the other hand, the graph-based ER approaches bring two benefits. First, it can capture the semantic relationships between different attributes within every tuple. Second, it can discover the rich set of semantics inherent among different tuples.
Recent studies~\cite{CreatingEmDI20,GraphER20} transform every dataset containing a collection of tuples into a graph composed of three types of nodes, i.e., \emph{tuple-level nodes}, \emph{attribute-level nodes}, and \emph{value-level nodes}.
The graph exhibits two characteristics:
(i) there is an edge between a tuple-level node and a value-level node as long as the value appears in the tuple; and
(ii) there is an edge between an attribute-level node and a value-level node if the value belongs to the domain of this attribute.
% To enrich the semantic features contained in the graph structure, GraphER~\cite{GraphER20} replaces the value-level nodes with token-level ones (a textual value can be split into multiple tokens) but keeps other types of nodes proposed by EMBDI.
However, graph-based ER is error-prone.
Take the sampled Amazon dataset in Figure~\ref{example:intro} as an example.
The value ``aspyr media inc'', which corresponds to a wrong attribute-level node, will result in a wrong graph structure.
The wrong graph features might be propagated along the edges and nodes, and thus lead to unreliable embeddings of tuples. Consequently, we are required to find sufficient tuple features in order to equip graph-based ER approaches with fault-tolerance.
%in a fault-tolerant way.
% Accordingly, if we absorb the advantages of both sentence-based methods and graph-based approaches, we can match tuples referring to the same real-world entity correctly.

% Cappuzzo et al.~\cite{CreatingEmDI20} has indicated that tuples are not sentences.

% it is not trivial to propose an explicit objective function to guide the training process of an embedding-based model effectively.
\vspace{0.05in}
\noindent
\textbf{Contributions.}
The obstruction with the existing ER methods inspires us to ask a question: would it be possible to perform ER in a \textbf{self-supervised} manner, where reliable labels are automatically generated and sufficient entities features are captured, so that the above two challenges could be well addressed?
Accordingly, we propose \textsf{CollaborER}, a self-supervised entity resolution, powered by multi-features collaboration.
\textsf{CollaborER} features a sequential modular architecture consisting of two phases, i.e., \emph{automatic label generation (ALG)} and \emph{collaborative ER training (CERT)}.
In the first phase, ALG is developed to generate reliable ER labels on every dataset automatically.
In the second phase, with the guidance of the generated labels, CERT learns the matching signals by utilizing both \emph{sentence features} and \emph{graph features} of tuples collaboratively.
% For capturing graph features, we first propose a \emph{multi-relational graph construction (MRGC)} approach to build a graph for each relational dataset and then learn the features/embeddings of tuples according to the graph structure.

We summarize the contributions of this paper as follows:
\begin{itemize}
    \item{\emph{Self-supervised ER framework.}}
    We propose a self-supervised ER framework \textsf{CollaborER}, which requires \textbf{zero} human involvement to generate labeled tuple pairs with high quality.
    Once the reliable labels are generated, \textsf{CollaborER} produces outstanding ER results via the collaboration of both \emph{sentence features} and \emph{graph features} of tuples.
    \item{\emph{Automatic label generation.}}
    We present ALG, for the first time, to automatically generate both \emph{positive} and \emph{negative} tuple pairs for the ER task.
    ALG greatly helps \textsf{CollaborER} to correctly identify ``challenging'' tuple pairs that are hard to tell whether they are matched.
    \item{\emph{Collaborative ER training.}}
    We introduce CERT, a collaborative ER training approach, to discover both \emph{graph features} and \emph{sentence features} to learn sufficient tuple features for ER without sacrificing the fault-tolerance capability in handling noisy tuples.
    \item{\emph{Extensive experiments.}}
    Comprehensive evaluation over eight existing ER benchmarks demonstrates the superiority of \textsf{CollaborER}. It outperforms all the existing unsupervised methods. Furthermore, it is comparable with or even superior to DITTO~\cite{Ditto20}, the state-of-the-art supervised ER method.
\end{itemize}

\noindent
\textbf{Organization.} The rest of the paper is organized as follows.
Section~\ref{sec:preliminaries} covers the basic background techniques used in the paper.
Section~\ref{sec:framework} presents the overall architecture of our proposed \textsf{CollaborER}, and Section~\ref{sec:alg} and Section~\ref{sec:cert} detail the two key phases of \textsf{CollaborER} respectively.
Section~\ref{sec:experiment} reports the experimental results and our findings.
Section~\ref{sec:demo} presents an anomaly detection demonstration based on the proposed \textsf{CollaborER}.
Section~\ref{sec:relatedwork} reviews the related work.
Finally, Section~\ref{sec:conclusions} concludes the paper.

\section{Preliminaries}
\label{sec:preliminaries}

In this section, we first formalize the problem of entity resolution and then overview some background materials and techniques to be used in subsequent sections. Table~\ref{tab:symbol} summarizes the symbols that are frequently used throughout this paper.

\begin{table}
\centering \small
\caption{Symbols and description}
\vspace*{-2mm}
\label{tab:symbol}
\setlength{\tabcolsep}{5mm}{
\begin{tabular}{|c|l|}
\hline
\textbf{Notation} & \textbf{Description} \\
\hline
$T$ & a relational dataset \\ \hline
$e\in T$ & a tuple belonging to the dataset $T$  \\ \hline
$A$ & a set of attribute values \\ \hline
$e.A[m]$ & the $m$-th attribute value of tuple $e$   \\ \hline
$\mathcal{G}$ & a multi-relational graph \\ \hline
$N$ & a set of nodes belonging to the graph $\mathcal{G}$ \\ \hline
$E$ & a set of edges belonging to the graph $\mathcal{G}$ \\ \hline
\end{tabular}}
\vspace*{-2mm}
\end{table}

% Bidirectional Encoder Representations from Transformers (BERT) [21], which is one of the state-of-the-art language pre-training models. BERT has achieved high quality results in many language understanding tasks (including text classification, question answering, etc.). In view of this, we adopt BERT as the underlying model for text classification.

\subsection{Problem Definition}
Let $T$ be a relational dataset with $|T|$ tuples and $m$ attributes $A = \{A[1], A[2], \cdots, A[m]\}$.
Each tuple $e \in T$ consists of a set of attribute values, denoted as $V = \{e.A[1], e.A[2], \cdots, e.A[m]\}$.
Here, $e.A[m]$ is the $m$-th attribute value of tuple $e$, corresponding to attribute $A[m] \in A$.
Entity resolution (ER), also known as entity matching or record linkage, aims to find matched tuple pairs $\mathcal{M}$ that refer to the same real-world entity between two relational datasets $T$ and $T'$.
%\baihua{What are $T$/$T'$? Collections of what (tuples)? $T$ and $T'$ share the same set of attributes? So they are two tables sharing the same table structures? }
The ER task can be formulated as $\mathcal{M} = \{(e, e') \in T \times T' | e \equiv e'\}$, where $e \in T$, $e' \in T'$, and $\equiv$ represents a matched relationship between tuples $e$ and $e'$.

To reduce the quadratic number of candidates of matched tuple pairs, an ER program often executes a \emph{blocking} phase followed by a \emph{matching} phase.
The goal of blocking is to identify a small subset of $T \times T'$ for candidate pairs of high probability to be matched.
In addition, blocking mechanisms are expected to have zero false negative, a common assumption made by many ER techniques~\cite{MudgalLRDPKDAR18, Ditto20, ZeroER20}.
Designing an effective blocking strategy is orthogonal to \textsf{CollaborER}, and we apply a common blocking method that is widely used in the existing ER approaches~\cite{Magellan16, DeepMatcher18, Ditto20}.
The goal of matching is to predict the matched tuple pairs in the candidate pairs, which is the focus of this work.

\subsection{Pre-trained Language Models}

Pre-trained language models (LMs), such as BERT~\cite{BERT19} and XLNet~\cite{XLNet19}, have demonstrated a powerful semantic expression ability.
Based on pre-trained LMs, we can support many downstream tasks (e.g., classification and question answering).
Concretely, we can plug appropriate inputs and outputs into a pre-trained LM based on the specific task and then fine-tune all the model's parameters end-to-end.

Intuitively, the ER problem can be treated as a sentence pair classification task~\cite{Ditto20}.
Given two tuples $e_i \in T$ and $e_j' \in T'$, pre-trained LMs transform them into two sentences $\mathcal{S}(e_i)$ and $\mathcal{S}(e_j')$, respectively.
A sentence $\mathcal{S}(e_i)$ is denoted by
$\mathcal{S}(e_i) ::= \langle$[COL] $A[1]$ [VAL] $e_i.A[1]$ ... [COL] $A[m]$ [VAL] $e_i.A[m]\rangle$,
where [COL] and [VAL] are special tokens for indicating the start of attribute names and the start of attribute values, respectively.
Note that, we exclude missing values and their corresponding attribute names from the sentence since they contain zero valid information.
A tuple pair $(e_i, e_j')$ can be serialized as a pairwise sentence
$\mathcal{S}(e_i, e_j') ::= \langle \mathcal{S}(e_i)$ [SEP] $\mathcal{S}(e_j')\rangle$, where [SEP] is a special token separating the two sentences.
For the sentence pair classification task, pre-trained LMs take each pairwise  sentence $\mathcal{S}(e_i, e_j')$ as an input. Note that, a special symbol [CLS] is added in front of every input sentence, which is utilized to store the classification output signals during the fine-tuning of LMs.

\noindent
\textbf{Objective Function.}
We employ \emph{CrossEntropy Loss}, a widely used classification objective function, to fine-tune the pre-trained LMs in \textsf{CollaborER}.
CrossEntropy Loss is designed to keep the predicted class labels similar to the ground-truth. Formally,
\begin{equation}\label{eq:crossentropy_loss}
    \mathcal{L}(y=k|\mathcal{S}(e_i, e_j'))  = -\log \left(\frac{\exp (d_k)}{\sum_{q}^{|k|} \exp (d_q)}\right)
    \forall k \in\{0,1\}
\end{equation}
Here, $\boldsymbol{d} \in \mathbb{R}^{|k|}$ is the logits computed by $\boldsymbol{d} = \mathbf{W}_{\rm c}^{\top} \mathbf{E_{[CLS]}}$.
$\mathbf{W}_c \in \mathbb{R}^{n\times|k|}$ is a learnable linear matrix, where $n$ is the dimension of the sentence embeddings.
$\mathbf{E_{[CLS]}}$ is the embedding of the symbol [CLS].
For sentence pair classification, the class labels are binary \{0, 1\}. We denote $y=1$ a truly matched pair and $y=0$ a mismatched pair.
%\baihua{Do we use $J$ somewhere? If not, no need to use this parameter, right?}\textcolor{red}{[cc: I've removed it.]}

\subsection{Graph Neural Networks}\label{sec:gnn}
Graph neural networks (GNNs) are popular  graph-based models, which capture graph features via message passing between the nodes of graphs.
GNNs are suitable for the ER task because of the following two aspects.
First, GNNs ignore the sequence relationship between different attributes but discover the features of each tuple by aggregating the semantic information contained in the corresponding attribute names and values. This conforms to the real characteristics of the relational dataset since a tuple is not a sentence and entities’ attributes can be organized in any order. Thus, GNNs are able to effectively capture the features within every tuple.
Second, recall that GNNs capture graph features via message passing between relevant nodes, i.e., the set of tuples sharing the same attribute values in this paper. Accordingly, GNNs have the capability of learning rich semantics among those relevant tuples, since the features of a tuple can be passed to another tuple through an edge (i.e., a shared attribute value).
% Recently, graph neural networks (GNNs) have shown success in various domains.
The core idea of GNNs is to learn each node representation by capturing the information passing from its neighborhoods.
% \footnote{Without loss of generality, we use the terms of “entity” and “tuple” interchangeably in the following experimental sections}
% \textcolor{red}{Many GNN-based methods \cite{AttrGNN20, MultiKE19, RREA20} have present promising performances by propagating the matching signal to the node’s neighbors.}
% Therefore, we incorporate GNN-based models in the proposed \textsf{CollaborER}.
Generally, GNNs learn the embeddings of each node $n_i$ obeying the following equations \cite{RREA20}:
\begin{gather}
\mathbf{o}_{i}^{l+1}=\text {AGGREGATION}^{l}\left(\left\{\!\!\left\{\left(\mathbf{h}_{j}^{l}, \mathbf{r}_{i, j}\right): j \in \mathcal{N}(i)\right\}\!\!\right\}\right)\\
\mathbf{h}_{i}^{l+1}=\text{UPDATE}^{l+1}\left(\mathbf{h}_{i}^{l}, \mathbf{o}_{i}^{l+1}\right)
\end{gather}
where $\mathbf{h}_{i}^{l}$ represents the embedding of the $l$-th layer node $n_i$, $\mathbf{r}_{i,j}$ stands for the embedding of an edge that connects the node $n_i$ and another node $n_j$, and $\{\!\!\left\{\cdots\right\}\!\!\}$ denotes a multiset.
$\mathcal{N}(i)$ represents the set of neighboring nodes around $e_i$.
Eq. (2) is to aggregate information from the neighboring nodes while Eq. (3) transforms the entity embeddings into better ones.
To serve the purpose of AGGREGATION, we can use graph convolutional network (GCN)~\cite{GCN17} or graph attention network (GAT)~\cite{GAT18}.

\section{Framework Overview}\label{sec:framework}

\begin{figure}[t]
\centering
\includegraphics[width=3.5in]{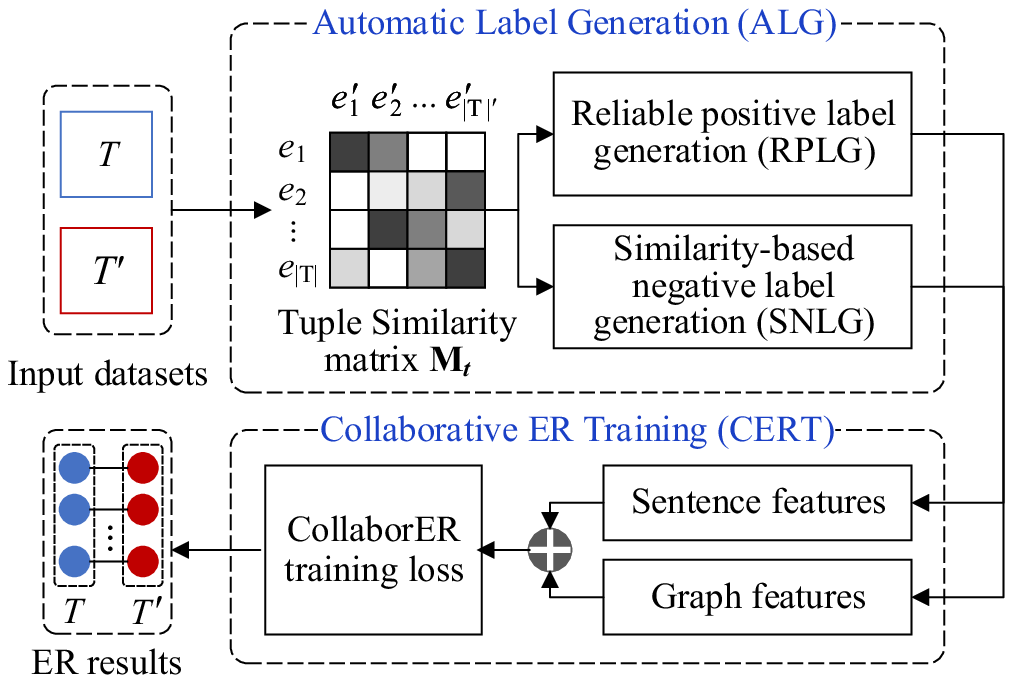}
\vspace{-5mm}
\caption{\textsf{CollaborER} framework}
\label{fig:framework}
\vspace{-2mm}
\end{figure}

In this section, we overview the framework of \textsf{CollaborER}, as illustrated in Figure~\ref{fig:framework}.
\textsf{CollaborER} consists of two phases, i.e., (i) automatic label generation (ALG), and (ii) collaborative ER training (CERT).

\vspace{0.05in}
\noindent
\textbf{Automatic label generation (ALG).}
As mentioned in Section~\ref{sec:introduction}, pre-collected ER labels are often not available in many real-world scenarios.
It inspires us to look for ways to generate approximate labels via an automatic label generation program.
Given two datasets $T$ and $T'$, each of which contains a collection of tuples, this phase is to generate pseudo-labels with high-quality, including both \emph{positive labels} and \emph{negative labels}, for the guidance of the subsequent CERT process.

Positive labels refer to a set of positive tuple pairs, denoted as $\mathbb{P}$.
For each positive tuple pair $\mathbb{P}(e_i, e_i')$, the tuple $e_i \in T$ and the tuple $e_i' \in T'$ have a high probability of being matched.
In ALG, we introduce a \emph{reliable positive label generation (RPLG)} strategy to obtain positive tuple pairs with high confidence.

On the other hand, negative labels refer to a set of negative tuple pairs, denoted as $\mathbb{N}$.
For each negative tuple pair $\mathbb{N}(e_i, e_j')$, the tuple $e_i \in T$ and the tuple $e_j' \in T'-\{e_i'\}$ are unlikely to be matched.
Random sampling~\cite{Mikolov13, Complex16} is a widely-used approach for generating negative labels.
Given a positive tuple pair $\mathbb{P}(e_i, e_i')$, random sampling replaces either $e_i$ or $e_i'$ with an arbitrary tuple.
However, recent studies~\cite{rotatE19, BootEA18} have indicated that the randomly generated negative tuple pairs are easily distinguished from positive ones.
For instance, if we generate a negative tuple pair (``Apple Inc.'', ``Google'') for a positive tuple pair (``Apple Inc.'', ``Apple''), it is obvious that ``Google'' and ``Apple Inc.'' are not equivalent.
These negative tuple pairs are uninformative, and contribute little to the embedding training process.
Ideally, an effective negative label generation is expected to put two similar tuples (but they are not related to the same real-world entity) into a pair.
It facilitates an ER-oriented embedding model (e.g., CERT in this paper) to be capable of identifying whether two entities of a ``challenging'' tuple pair refer to the same real-world entity.
To this end, we propose a \emph{similarity-based negative label generation (SNLG)} method in ALG to generate negative labels with semantic similarity.

\vspace{0.05in}
\noindent
\textbf{Collaborative ER training (CERT).}
Recall that matching entities purely based on the \emph{sentence features} or the \emph{graph features} of tuples results in insufficient feature discovery or erroneous feature involvement.
The goal of CERT is to capture and integrate %collaborate
both the sentence features and the graph features of tuples in a unified framework to improve the quality of ER results.

Given two datasets $T$ and $T'$ and a set of labels (including positive labels $\mathbb{P}$ and negative labels $\mathbb{N}$) generated by ALG, CERT first introduces \emph{multi-relational graph construction (MRGC)} to construct a multi-relational graph $\mathcal{G}$ (w.r.t. $\mathcal{G}'$) for each dataset $T$ (w.r.t. $T'$).
We would like to highlight that the graph structure generated by the proposed MRGC is much \textbf{simpler} than that generated by other existing ER methods (e.g., EMBDI~\cite{CreatingEmDI20} and GraphER~\cite{GraphER20}) without losing the expressive power of tuples' graph features, as confirmed in the experimental evaluations to be presented in Section~\ref{exp:HGC_analysis}.

Then, CERT learns the embeddings of each tuple based on the graph structure.
CERT is treated as a black box, such that users could enjoy the flexibility of applying their choice of graph-based models to embed both nodes and edges in a multi-relational graph.
Our current implementation utilizes AttrGNN~\cite{AttrGNN20} for this purpose.
Afterward, we feed the well-trained graph features (i.e., embeddings) of tuples into a pre-trained language model (LM) to assist the learning of the sentence features of tuples.
More specifically, the graph features of tuples are used to complement the semantic features of tuples that cannot be captured by a sentence-based model.

\section{Automatic Label Generation (ALG)}
\label{sec:alg}

In this section, we present an automatic label generation (ALG) strategy.
It contains two components, including (i) a reliable positive label generation (RPLG) method and (ii) a similarity-based negative label generation (SNLG) method.

Generating either positive labels or negative labels is highly relevant to the similarity between tuples.
Motivated by the powerful capability of semantics expression of pre-trained language models, we leverage sentence-BERT~\cite{sentenceBERT19}, a variant of BERT that achieves outstanding performance for semantic similarity search, to assign a pre-trained embedding for each tuple.
In general, different similarity functions (e.g., cosine distance and Manhattan distance) can be applied to quantify the semantic similarity between tuples from different datasets in ALG, according to the characteristics of the datasets.
In the current implementation, we find empirically that cosine distance brings considerable performance. To this end, we choose it as the similarity function in ALG.
The tuple similarity matrix is denoted as $\mathbf{M}_t \in [0,1]^{|T| \times |T'|}$, where $|T|$ and $|T'|$ represent the total number of tuples in $T$ and $T'$ respectively. In the following, we detail how to generate positive and negative labels via RPLG and SNLG, respectively.

%\vspace{0.05in}
%\noindent
%\textbf
\subsection{Reliable Positive Label Generation (RPLG)}

\begin{figure*}[t]
\centering
\includegraphics[width=7.3in]{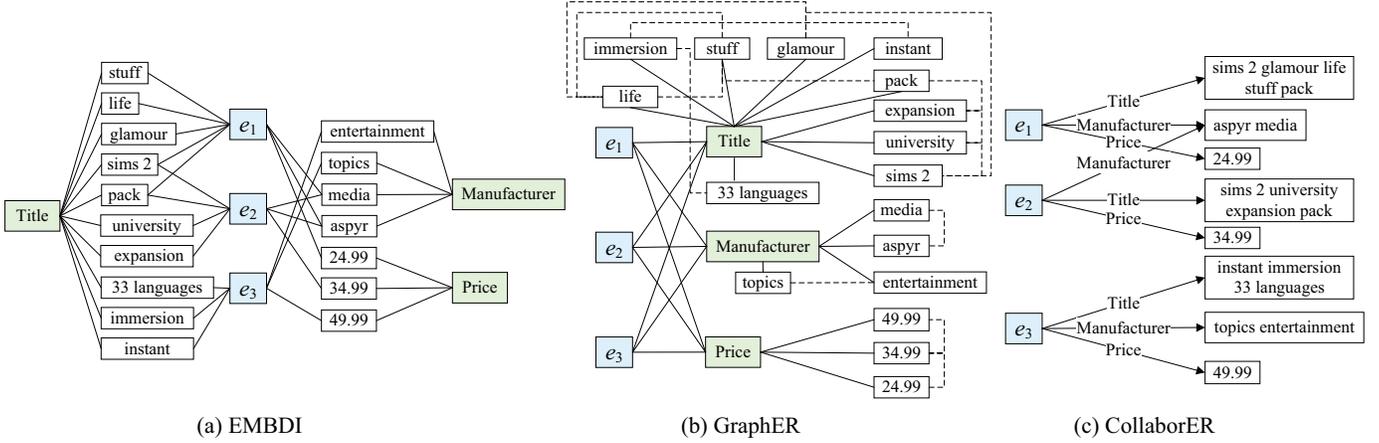}
\vspace{-6mm}
\caption{A motivating example of proposing the multi-relational graph construction (MRGC)}
\label{fig:table2kg}
\vspace*{-2mm}
\end{figure*}

RPLG aims to find positive tuple pairs with a high probability of being matched.
%Intuitively, if two tuples are matched, they are mutually the most similar to each other.
A common approach is to consider tuples that are mutually most similar to each other.
However, we find empirically many mutually similar tuples do not refer to the same entities.
Considering that high-quality labels are essential for embedding-based ER model as wrong labels will mislead the ER model training, we choose to generate the positive labels by IKGC~\cite{DAT20}, which gives much stronger constraints to ensure the high-quality of positive labels than the methods based on the mutual similarity.
It generates tuple pairs as positive labels that satisfy two requirements~\cite{DAT20}, including (i) they are mutually most similar to each other; and (ii) there is a margin between, for each tuple $e$, its most similar tuple and the second most similar one.
%
%(i) the similarity between the two tuples, which belong to the same positive label, is the highest from both sides; and (ii) there exists a margin between the top-2 candidates.

Specifically, for each tuple $e_i \in T$, we assume that $e_j', e_k' \in T'$ are the most similar and the second most similar tuples in $T'$ to $e_i$, respectively.
%We use $Sim(e_i, e_j')$ to represent the similarity score between $e_i$ and $e_j'$ and use $Sim(e_i, e_k')$ to denote the similarity score between $e_i$ and $e_k'$.
Similarly, for tuple $e_j' \in T'$ (i.e., the most similar tuple to $e_i \in T$), let $e_l$ and $e_u$ denote its most similar tuple in $T$ and the second most similar tuple in $T$ respectively.
%
%, we can compute the similarity score between $e_j'$ and its most similar tuple $e_l$ (w.r.t. is second most similar tuple $e_u$), denoted as $Sim(e_j', e_l)$ (w.r.t. $Sim(e_j', e_u)$).
%
If tuple pair $(e_i, e_j')$ could be considered as a positive label, we expect $e_i = e_l$, i.e., $e_i$ and $e_j'$ are mutually most similar to each, i.e., the requirement (i) stated above.
%If $e_i = e_l$ (i.e., they are the same tuple) and
In addition, their similarity discrepancies $\delta_1 = Sim(e_i, e_j') - Sim(e_i, e_k')$ and $\delta_2 = Sim(e_j', e_l) - Sim(e_j', e_u)$ are expected to be both above a given threshold $\theta$, i.e., the requirement (ii) stated above. Here, $Sim(e, e')$ denotes the similarity score between two tuples $e \in T$ and $e'\in T'$.
%, the tuple pair $(e_i, e_j')$ would be considered as a positive label.

\vspace{0.05in}
\noindent
\textbf{Discussion.}
We would like to emphasize that RPLG is a general approach, which can be easily integrated into various ER methods.
RPLG is able to not only generate positive labels with high-quality (see the experiments to be presented in Section~\ref{exp:ALG_analysis}) but also achieve desirable ER results without any time-consuming training process (see the experiments to be presented in Section~\ref{sec:ablation}).

%\vspace{0.05in}
%\noindent
%\textbf
\subsection{Similarity-based Negative Label Generation (SNLG)}

As the random-based negative label generation method has rather limited contribution to the embedding-based ER training, it is essential to generate more ``challenging'' negative labels, as described in Section~\ref{sec:framework}.
To achieve this goal, we propose a similarity-based negative label generation (SNLG) strategy.
Given a positive tuple pair $\mathbb{P}(e_i, e_i')$, where $e_i \in T$ and $e_i' \in T'$, SNLG generates a set of negative labels $\mathbb{N}(e_i, e_i')$ by replacing either $e_i$ or $e_i'$ with its $\epsilon$-nearest neighborhood in the semantic embedding space.
Again, we use the cosine similarity metric to search for the $\epsilon$-nearest neighbors of $e_i$ and $e_i'$, respectively.

\vspace{0.05in}
\noindent
\textbf{Discussion.}
Even though this is a very intuitive and simple method, it effectively promotes the performance of \textsf{CollaborER}.
We will demonstrate the superiority of using the proposed SNLG to generate negative labels for ER in the experiments to be presented in Section~\ref{sec:ablation}.

\section{Collaborative ER Training (CERT)}
\label{sec:cert}

This section details a newly proposed collaborative ER training (CERT) approach to discover the features of tuples from both the graph aspect and the sentence aspect to facilitate the ER process.
CERT is composed of two phases, i.e., (i) multi-relational graph feature learning (MRGFL) and (ii) collaborative sentence feature learning (CSFL).

\subsection{Multi-Relational Graph Feature Learning (MRGFL)}

Inspired by the graph structure's powerful capturing ability of semantics, we propose a multi-relational graph feature learning method (MRGFL) to represent tuples according to their graph features.
It first proposes a \emph{multi-relational graph construction} (MRGC) approach for transforming datasets from the relational format to the graph structure, and it then learns the tuple representations via a GNN-based model, e.g., AttrGNN~\cite{AttrGNN20} in our current implementation.

\begin{algorithm}[t]
\LinesNumbered
\DontPrintSemicolon
\caption{Multi-Relational Graph Construction}
\label{algorithm:graph_construction}
    \KwIn{a relational dataset $T$}
    \KwOut{a multi-relational graph $\mathcal{G}$}
    $\mathcal{G} \longleftarrow \varnothing$\;
    \ForEach{$e_i \in T$}{
        $\mathcal{G}$.addNode($e_i$)\;
        \ForEach{$v_j \in \{e_i.A[1], e_i.A[2], \cdots, e_i.A[m]\}$}{
            \If{$v_j$ is not included in $\mathcal{G}$}{
                $\mathcal{G}$.addNode($v_j$)\;
            }
            $a_{i,j} \longleftarrow$ find the attribute name of $v_j$\;
            $\mathcal{G}$.addEdge($e_i, a_{i,j}, v_j$)\;
        }
    }
\Return{$\mathcal{G}$}
\end{algorithm}

\vspace{0.05in}
\noindent
\textbf{Multi-relational graph construction (MRGC).}
Graph construction techniques have been presented in the existing ER work, such as EMBDI~\cite{CreatingEmDI20} and GraphER~\cite{GraphER20}.
These techniques treat tuples, attribute values, and attribute names as three different types of nodes. Edges exist if there are relationships between nodes.
Nonetheless, several drawbacks restrict the scope of using these graph construction methods to perform ER in real-world scenarios.

First, these graph construction approaches may produce intricately large-scale graphs containing a large number of edges and nodes.
Storing a graph with massive edges and nodes is memory-consuming, and meanwhile training a graph embedding model (e.g., GNN) on a large graph is challenging too, as widely-acknowledged by many existing works~\cite{distDGL20, HamiltonYL17, ChenZS18}.

Second, these graph construction methods lack consideration of the semantics contained in an edge itself.
For instance, assume that there are two types of edges, i.e., \emph{attribute-value edges} and \emph{tuple-attribute edges}.
The former represents an edge that connects an attribute-level node with a value-level node; the latter represents an edge connecting a tuple-level node with an attribute-level node.
%a node with the attribute name type and a node with the attribute value type. The latter represents an edge that connects a node with the tuple type and a node with the attribute name type.
%
It is intuitive that these two types of edges have different semantic meanings, and hence, they should be considered differently when learning features of tuples in the ER task.

The above limitations motivate us to design a relatively \textbf{small-scale} but highly \textbf{effective} MRGC to construct a multi-relational graph for every dataset.
We start by defining a multi-relational graph, formally $\mathcal{G} = \{N, E, A\}$. Here, $N$ and $E$ refer to a set of nodes and a set of edges, respectively; and $A$ represents the set of attributes corresponding to the nodes and the edges.
There are two types of nodes in $\mathcal{G}$, i.e., \emph{tuple-level nodes} and \emph{value-level nodes}.
A tuple-level node represents a tuple $e$; while a value-level node corresponds to an attribute value $v$ in a relational dataset.
Each attribute $a \in A$ denotes an attribute name in the relational dataset.
%\baihua{Is relation a commonly used term? Relation refers to a table in relational databases. Why not use attribute instead?}
${E} = \{(e, a, v) | e, v \in N, a \in A\}$ represents the set of edges.
%, each of which connects a tuple and one of its relevant attribute value.
%
Each edge connects a tuple-level node $e$ with a value-level node $v$ via an attribute $a$, meaning that $e$ has $v$ as its value for attribute $a$.

Next, we describe the MRGC procedure, with its pseudo code presented in Algorithm~\ref{algorithm:graph_construction}.
Given a relational dataset $T$, MRGC initializes an empty multi-relational graph $\mathcal{G}$ (Line 1).
Then, MRGC iteratively adds nodes and edges to $\mathcal{G}$ (Lines 2-8).
Specifically,
for every tuple $e_i\in T$, MRGC first selects its tuple Id as a tuple-level node (Line 3) and then adds a set of value-level nodes that correspond to this tuple (Lines 4-6). Note that, since different tuples share the same attribute names, MRGC generates a set of edges for $e_i$, with each connecting the tuple-level node of $e_i$ and a value-level node $v_j \in \{e_i.A[1], e_i.A[2],...,e_i.A[m]\}$, denoted as $(e_i, a_{i,j}, v_j)$.

\vspace{0.05in}
\noindent
\textbf{Discussion.}
Compared to the existing graph construction methods, MRGC constructs a small graph that is still able to well preserve the semantics of tuples.
Take the sampled Amazon dataset as an example.
Figure~\ref{fig:table2kg} shows the respective graph structures constructed by three different graph construction methods, including EMBDI~\cite{CreatingEmDI20}, GraphER~\cite{GraphER20}, and the proposed MRGC in this paper.
It is obvious that the graph constructed by MRGC is the smallest, containing fewer nodes and edges than other graphs.
Also, we will verify the small-scale characteristics of MRGC in the experiments to be presented in Section~\ref{exp:HGC_analysis}.
Besides, MRGC not only preserves the semantic relationships between each tuple and its corresponding attribute values, but also maintains semantic connections between different tuples by connecting them with a shared value-level node.
For example, $e_1$ and $e_2$ have semantic connections since they both have edges linking to the same value-level node ``aspyr media''.

\vspace{0.05in}
\noindent
\textbf{Multi-relational graph feature learning (MRGFL).}
% Recall that GNNs have demonstrated the effectiveness of learning the node embeddings by capturing the graph features, as illustrated in Section~\ref{sec:gnn}.
% In the following, we illustrate how to learn tuples' graph features in the ER task.
%
Given two multi-relational graphs $\mathcal{G}$ (w.r.t. $T$) and $\mathcal{G}'$ (w.r.t. $T'$), MRGFL aims to embed tuples from different sources in a unified vector space by considering their graph structures.
In this vector space, matched tuples are expected to be as close to each other as possible.
Generally, we can treat MRGFL as a graph-based ER problem, which is highly relevant to \emph{entity alignment} (EA)~\cite{OpenEA2020VLDB} that aims to find a correspondence between entities from different multi-relational graphs.
To this end, MRGFL is regarded as a black box.
Users have the flexibility to learn the embeddings of tuples by applying any available EA model~\cite{OpenEA2020VLDB, AttrGNN20, KECG19}.
%Since EA models have been extensively studied~\cite{OpenEA2020VLDB, AttrGNN20, KECG19},
In our implementation, we adopt the state-of-the-art EA model AttrGNN~\cite{AttrGNN20} that aggregates the graph feature of each tuple via multiple newly proposed GNNs, for this purpose.
In the following, we sketch the main idea about how to use an EA model to learn tuples' graph features in MRGFL.

First, the graph features of each tuple can be obtained by applying a GNN model, as described in Section~\ref{sec:gnn}. It outputs a set of tuples' embeddings. We denote the embedding of each tuple $e_i \in T$ (w.r.t. $e_i' \in T'$) as $\mathbf{h}_{e_i}$ (w.r.t. $\mathbf{h}_{e_i'}$).
Then, a \emph{training objective function} (denoted as $\mathcal{L}_{g}$) is used to unify the two datasets' tuple embeddings into a unified vector space by maximizing the similarities with regard to the generated labels.
Formally,
\begin{equation}
    \mathcal{L}_{g}=\sum_{(e_i, e_i')  \in \mathbb{P}} \sum_{(e_j, e_k)  \in \mathbb{N}} \left[d(e_i, e_i') + \gamma - d(e_j, e_k) \right]_{+}
\end{equation}
Here, $(e_i, e_i') \in \mathbb{P}$ represents a positive label;
$(e_j, e_k) \in \mathbb{N}$ represents a negative label;
$[b]_{+} = max\{0, b\}$;  $d(e_i, e_i')$ denotes the cosine distance between $\mathbf{h}_{e_i}$ and $\mathbf{h}_{e_i'}$, where $\mathbf{h}_{e_i}$ and $\mathbf{h}_{e_i'}$ are the final embeddings of $e_i$ and $e_i'$ w.r.t. the multi-relational graph $\mathcal{G}$ after performing the $|l|$-th layer GNN model, respectively;
similarly, $d(e_j, e_k)$ represents the cosine distance between $\mathbf{h}_{e_j}$ and $\mathbf{h}_{e_k}$;
and
% \baihua{Confused by the above sentence. Isn't $(e_i, e_i') \in \mathbb{P}$? Do you mean the negative tuple pairs generated based on $(e_i, e_i') \in \mathbb{P}$?}
$\gamma$ is a margin hyper-parameter.
We set $\gamma = 1.0$ in the current implementation.

\subsection{Collaborative Sentence Feature Learning (CSFL)}

\begin{figure}[t]
\centering
\includegraphics[width=3.65in]{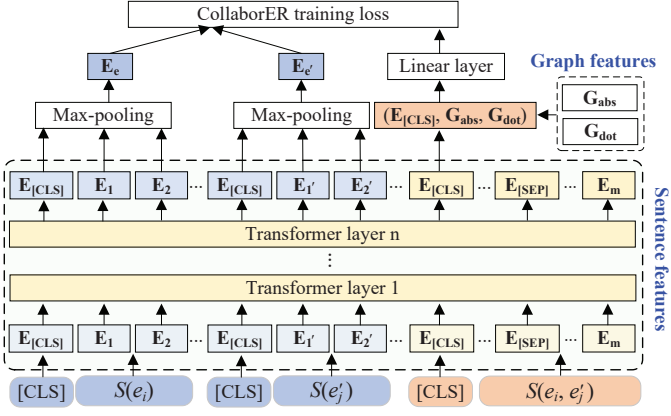}
\vspace{-6mm}
\caption{The architecture of CSFL}
\label{fig:CSFL_model}
\vspace{-3mm}
\end{figure}

Recent studies~\cite{Ditto20, AutoEM19, DeepER18} have demonstrated that capturing the sentence features of tuples can help ER task to a certain degree.
Nonetheless, treating tuples as sentences causes insufficient feature discovery, as mentioned in Section~\ref{sec:introduction}.
In view of this, we propose a collaborative sentence feature learning (CSFL) model, which discovers sufficient tuples' sentence features for ER with the assistance of the well-trained graph features of tuples.
The training objective of CSFL is to (i) identify whether two tuples refer to the same real-world entity; and (ii) minimize the semantic distance between the matched tuples. The architecture of CSFL is depicted in Figure~\ref{fig:CSFL_model}.

First, we present how to identify the matched (or mismatched) tuples in CSFL.
We fine-tune a pre-trained LM with a sentence pair classification task.
We take as inputs a pairwise sentence $\mathcal{S}(e_i, e_i')$ and its corresponding positive and negative labels generated by the proposed ALG.
Then, we learn the classification signal $\rm{\mathbf{E_{[CLS]}}}$ by feeding the inputs into a multi-layer Transformer encoder.
In the current implementation, the number of transformer layers is set to 12, a typical setting used in various tasks.
We use a variant of \emph{CrossEntropy Loss} $\mathcal{L}_1$ as the objective training function, which is derived from Equation~(\ref{eq:crossentropy_loss}). Formally,
\begin{gather}
\mathcal{L}_1(y=k|\mathcal{S}(e_i, e_j'))  = -\log \left(\frac{\exp (d_k^{*})}{\sum_{q}^{|k|} \exp (d_q^{*})}\right) \forall k \in\{0,1\}\\
\boldsymbol{d}^{*} = \mathbf{W}_{\rm c}^{*\!\top}\!\! \left(\mathbf{E_{[CLS]}} ; \mathbf{G_{abs}} ; \rm{\mathbf{G_{dot}}} \right)
\end{gather}
Here,
the logits $\boldsymbol{d}^{*}$ is produced by both tuples' sentence features (i.e., $\mathbf{E_{[CLS]}}$) and tuples' graph features (i.e., $\mathbf{G_{abs}} \in \mathbb{R}^{c}$ and $\mathbf{G_{dot}} \in \mathbb{R}^{c}$), where $c$ is the dimension of the tuples' graph features.
$\mathbf{W}_{\rm c}^{*} \in \mathbb{R}^{(n+2c)\times |k|}$.
$\mathbf{G_{abs}} = |\mathbf{h}_{e_i} - \mathbf{h}_{e_i'}|$ denotes the element-wise difference between the graph-based embeddings $\mathbf{h}_{e_i}$ and $\mathbf{h}_{e_i'}$.
$\mathbf{G_{dot}} = \mathbf{h}_{e_i} \otimes \mathbf{h}_{e_i'}$ represents the element-wise similarity between $\mathbf{h}_{e_i}$ and $\mathbf{h}_{e_i'}$.

Second, we illustrate how to minimize the semantic distance between the matched tuples.
At the input, the pre-trained LM allocates an initialized embedding for each token of a sentence $\mathcal{S}(e_i)$ (w.r.t. $\mathcal{S}(e_i')$), denoted as $\rm{\mathbf{E_i}}$ (w.r.t. $\rm{\mathbf{E_i'}}$).
Note that, the special symbol [CLS], which is located in the front of every sentence, also has an initial embedding, denoted as $\rm{\mathbf{E_{[CLS]}}}$.
The embedding of every token will be updated after performing the multi-layer Transformer encoder.
We apply \emph{max-pooling} to obtain a fixed-length embedding $\rm{\mathbf{E_{e_i}}}$ (w.r.t. $\rm{\mathbf{E_{e_i'}}}$) for representing the tuple $e$ (w.r.t. $e'$).
Concretely, max-pooling generates the fixed-length embedding by selecting the maximal value in each dimension among all the embedded tokens of the tuple.
We use \emph{CosineEmbedding Loss} $\mathcal{L}_2$ as the objective training function. It is designed to minimize the semantic distance between matched tuples (w.r.t. the set of positive labels $\mathbb{P}$) and maximize that between mismatched tuples (w.r.t. the set of negative labels $\mathbb{N}$). Formally,
\begin{equation}\label{eq:cosineEmbeddingLoss}
    \mathcal{L}_2(y|\mathcal{S}(e_i), \mathcal{S}(e_j'))=\left\{\begin{array}{ll}
1- \cos (\mathbf{E_{e_i}}, \mathbf{E_{e_j'}}), \text { if } y=1 \\
\max (0, \cos (\mathbf{E_{e_i}}, \mathbf{E_{e_j'}})-\lambda), \text { if } y=0
\end{array}\right.
\end{equation}
where $\lambda$ is a margin hyper-parameter separating matched tuple pairs from mismatched tuple pairs. $\cos (\cdot,\cdot)$ represents the cosine distance metric.

Finally, we are ready to present the overall training function of CSFL, namely, \emph{CollaborER training loss} (denoted as $\mathcal{L}_{c}$). Formally,
\begin{equation}
\mathcal{L}_{c} = \mathcal{L}_1 + \mu \mathcal{L}_2
\end{equation}
where hyper-parameter $\mu \in [0,1]$ is a coefficient controlling the relative weight of $\mathcal{L}_2$ against $\mathcal{L}_1$.
The ER results can be obtained according to the predicted labels of each tuple pair.
% First, we describe how CSFL obtains sentence features of tuples.
% Inspired by the success of pre-trained LMs in the ER task, we fine-tune a pre-trained LM for this purpose.
% Given two tuples $e \in T$ and $e' \in T'$, we take three sentences as input, including (i) two serialized tuple $\mathcal{S}(e)$ and $\mathcal{S}(e')$ derived from $e$ and $e'$ respectively; and (ii) one pairwise sentence $\mathcal{S}(e, e')$.

\vspace{0.05in}
\noindent
\textbf{Discussion.}
Compared to the existing sentence-based ER methods that also fine-tune pre-trained LMs, we emphasize the superiority of CSFL in the following two aspects.
First, CSFL incorporates the graph features of tuples learned in the previous MRGFL step to enrich the features that the sentence-based model fails to capture.
Second, we argue that utilizing the CosineEmbedding Loss (i.e., $\mathcal{L}_2$ defined in Equation~\ref{eq:cosineEmbeddingLoss}) as a part of \emph{CollaborER training loss} is suitable for the ER task.
Intuitively, matched tuples should have similar embeddings in a unified semantic vector space.
However, the existing sentence-based ER methods, which fine-tune and cast ER
as a sentence-pair classification problem, cannot ensure the semantic similarity between matched tuples.
We will verify the superiority of the proposed CSFL in the experiments to be presented in Section~\ref{sec:ablation}.

\section{Experiments}
\label{sec:experiment}

% \begin{table}[t]
% \caption{Statistics of the datasets used in experiments}
% \setlength{\tabcolsep}{0.6mm}{
% \begin{tabular}{c|c|c|ccc}
% \toprule
% \textbf{Type} & \textbf{Dataset} & \textbf{Domain} & \textbf{Size} & \textbf{\# Pos.} & \textbf{\# Attr.} \\ \hline
% \multirow{5}{*}{Structured} & AG, Amazon-Google & software & 11,460 & 1,167 & 3 \\
%  & BR, BeerAdvo-RateBeer & beer & 450 & 68 & 4 \\
%  & DA-clean, DBLP-ACM & citation & 12,363 & 2,220 & 4 \\
%  & FZ, Fodors-Zagats & restaurant & 946 & 110 & 6 \\
%  & IA-clean, iTunes-Amazon & music & 539 & 132 & 8 \\ \hline
% \multirow{2}{*}{Dirty} & DA-dirty, DBLP-ACM & citation & 12,363 & 2,220 & 4 \\
%  & IA-dirty, iTunes-Amazon & music & 539 & 132 & 8 \\ \hline
% Textual & AB, Abt-Buy & product & 9,575 & 1,028 & 3 \\
% \bottomrule
% \end{tabular}}
% \label{tb:datasets}
% \end{table}

In this section, we conduct comprehensive experiments to verify the performance of \textsf{CollaborER} from the following three aspects.
First, we compare \textsf{CollaborER} with several competing ER approaches and present the results in Section~\ref{exp:overall}.
Second, we conduct the ablation study for the proposed \textsf{CollaborER} and report our findings in Section~\ref{sec:ablation}.
Third, we further explore \textsf{CollaborER} by (i) comparing the scale of the graphs generated by the proposed multi-relational graph construction (MRGC) method and other existing approaches in Section~\ref{exp:HGC_analysis}; and (ii) analyzing the performance of both the reliable positive label generation (RPLG) and the similarity-based negative label generation (SNLG) (in the automatic label generation (ALG) strategy) in Section~\ref{exp:ALG_analysis}.
%Last, we develop a visualized cross-platform system for anomaly detection based on \textsf{CollaborER}.

% (iii) investigating the effectiveness and efficiency of \textsf{CollaborER} for large-scale ER task;

\subsection{Benchmark Datasets}

We conduct experiments on \emph{eight} representative and widely-used ER benchmarks with different sizes and in various domains.
Table~\ref{tb:datasets} lists the detailed statistics.
For structured ER, we use five benchmarks, including Amazon-Google (AG), BeerAdvo-RateBeer (BR), the clean version of DBLP-ACM (DA-clean), Fodors-Zagats (FZ), and the clean version of iTunes-Amazon (IA-clean).
The attribute values of tuples are atomic but not a composition of multiple values.
For dirty ER, following~\cite{DeepMatcher18}, we use the dirty versions of the DBLP-ACM and iTunes-Amazon benchmarks to measure the robustness of the proposed \textsf{CollaborER} against noise.
For textual ER, we use the Abt-Buy (AB) benchmark which is text-heavy, meaning that at least one attribute of each tuple contains long textual values.
% \baihua{one attribute of each tuple?}\textcolor{red}{[cc: I've revised this part.]}

\begin{table}[t]
\caption{Statistics of the datasets used in experiments}
\vspace{-2mm}
\setlength{\tabcolsep}{0.55mm}{
\begin{tabular}{c|c|c|cccc}
\toprule
\textbf{Type}               & \textbf{Dataset}         & \textbf{Domain} & \textbf{\#Attr.} & \textbf{\#Domain} & \textbf{\#Tuple} & \textbf{\#Pos.} \\ \hline
\multirow{5}{*}{Structured} & AG        & software    & 3   & 123 - 3,021 & 1,363 - 3,226    & 1,167           \\
                            & BR    & beer     & 4   & 4 - 4,343    & 4,345 - 3,000    & 68              \\
                            & DA-clean      & citation & 4   & 5 - 2,507    & 2,616 - 2,294    & 2,220           \\
                            & FZ        & restaurant  & 6  & 16 - 533  & 533 - 331        & 110             \\
                            & IA-clean & music     & 8  & 6 - 38,794    & 6,907 - 55,923   & 132             \\ \hline
\multirow{2}{*}{Dirty}      & DA-dirty      & citation   & 4  & 6 - 2,588   & 2,616 - 2,294    & 2,220           \\
                            & IA-dirty & music     & 8  & 7 - 55,727    & 6,907 - 55,923   & 132             \\ \hline
Textual                     & AB    & product  & 3   & 167 - 1,081    & 1,081 - 1,092    & 1,028           \\ \bottomrule
\end{tabular}}
\label{tb:datasets}
\vspace{-2mm}
\end{table}

\begin{table*}[t]
\caption{Overall ER results with and without any pre-defined labels (F1-score values are in percentage, and the best scores are in \textbf{bold})}
\vspace*{-1mm}
\begin{threeparttable}
\setlength{\tabcolsep}{1.5mm}{
\begin{tabular}{c|c|cc|c||ccccc|cc}
\toprule
\multirow{2}{*}{\textbf{Type}}    & \multirow{2}{*}{\textbf{Datasets}} & \multicolumn{2}{c|}{\textbf{Unsupervised}} & \textbf{Self-supervised} & \multicolumn{5}{c|}{\textbf{Supervised}}     & \multicolumn{2}{c}{\textbf{Self-supervised}} \\ \cline{3-12}
       &        & \textbf{ZeroER*}     & \textbf{EMBDI}    & \textbf{CollaborER-U} & \textbf{DM+}    & \textbf{GraphER} & \textbf{MCA}  & \textbf{ERGAN}  & \textbf{DITTO}   & \textbf{DITTO-S} & \textbf{CollaborER-S} \\ \hline
\multirow{5}{*}{Structured} & AG        & 48.00      & 59.00    & \textbf{68.61}  & 70.70  & 68.08   & 71.40 & 37.49 & \textbf{75.58}  & 65.01 & 71.91     \\
       & BR        & 51.50      & 86.00    & \textbf{87.69}  & 78.80  & 79.71   & 80.00 & 74.42 & 94.37  & 90.32 & \textbf{96.55}  \\
       & DA-clean    & 96.00      & 95.00    & \textbf{98.63}  & 98.45  & --   & 98.90 & 98.51 & \textbf{98.99}  & 98.33 & 98.65     \\
       & FZ        & \textbf{100}  & 99.00    & \textbf{100} & \textbf{100} & --   & -- & 98.48 & \textbf{100}   & \textbf{100} & \textbf{100} \\
       & IA-clean    & 82.40      & 11.00    & \textbf{96.12}  & 91.20  & --   & -- & 77.29 & 97.06  & 94.12  & \textbf{100} \\ \hline
\multirow{2}{*}{Dirty}   & DA-dirty    & 63.00      & --    & \textbf{98.25}  & 98.10  & --   & 98.50 & 81.79 & 99.03  & 98.87  & \textbf{99.10}  \\
       & IA-dirty    & /     & --    & \textbf{95.17}  & 79.40  & --   & -- & 67.11 & 95.65  & 83.02  & \textbf{98.18}  \\ \hline
Textual      & AB        & 52.00      & 82.50    & \textbf{83.17}  & 62.80  & --   & 70.80 & 30.37 & \textbf{89.33} & 78.49 & 85.01     \\ \bottomrule
\end{tabular}}\label{tb:overall_ER_results}
\begin{tablenotes}
\footnotesize
 \item[1] The symbol ``*''  represents the corresponding ER results obtained by our re-implementation with the publicly available source code.
 \item[2] The symbol ``/'' indicates that the ER model \textbf{fails} to produce any result after running for 5 days in the experimental conditions.
 \item[3] The symbol ``--'' denotes that the results are not provided in the original paper.
\end{tablenotes}
\end{threeparttable}
\end{table*}

\subsection{Implementation and Experimental setup}

\vspace{0.05in}
\noindent
\textbf{Evaluation metric.}
To measure the quality of ER results, we use \emph{F1-score}, the harmonic mean of precision (\emph{Prec.}) and recall (\emph{Rec.}) computed as $\frac{2 \times (Prec. \times Rec.) }{ (Prec. + Rec.)}$.
Here, precision is defined as the fraction of match predictions that are correct;
and recall is defined as the fraction of real matches being predicted as matches.

\vspace{0.05in}
\noindent
\textbf{Competitors.}
We compare \textsf{CollaborER} against 6 SOTA ER approaches.
The competitors can be classified into two categories based on whether pre-defined lables are required, i.e., \emph{unsupervised ER} and \emph{supervised ER}.

The former refers to the group of approaches that performs ER without any label involvement, including
(i) ZeroER~\cite{ZeroER20}, a powerful generative ER approach based on Gaussian Mixture Models for learning the match and unmatch distributions; and
(ii) EMBDI~\cite{CreatingEmDI20}, which automatically learns local embeddings of tuples for ER based on the attribute-centric graphs.
Methods in this group are most relevant to \textsf{CollaborER}.

The latter refers to the group of approaches that relies on the pre-defined labels for matching tuples, including
(i) DeepMatcher+ (DM+)~\cite{Ditto20}, which implements multiple ER methods and reports the best performance (highest F1-scores), including DeepER~\cite{DeepER18}, Magellan~\cite{Magellan16}, DeepMatcher~\cite{DeepMatcher18}, and DeepMatcher's follow-up work~\cite{ERIJCAI19} and \cite{ActiveTransferER19};
(ii) GraphER~\cite{GraphER20}, which integrates schematic and structural information into token representations with a GNN model for ER and aggregates token-level features as the ER results;
%The ER results are fetched by aggregating token-level features;
(iii) MCA~\cite{MCA20}, which incorporates attention mechanism into a sequence-based model to learn features of tuples for ER;
(iv) ERGAN~\cite{ERGAN}, which employs a generative adversarial network to augment labels and predict whether two entities are matched;
and (v) DITTO~\cite{Ditto20}, which leverages a pre-trained Transformer-based language model to fine-tune and cast ER as a sentence-pair classification problem.
Approaches in this group are used to demonstrate that the proposed \textsf{CollaborER}, although not requiring any labor-intensive annotations/labels, is able to achieve performance that is comparable with or even better than the performance achieved by SOTA supervised ER in various real-world ER scenarios.
%without any labor-intensive annotations/labels.

Note that, in the evaluation of supervised ER methods, each dataset is split into the training, validation, and test sets using the ratio of 3:1:1.
For fair comparisons with supervised methods, we report the results conducted by \textsf{CollaborER} on the test sets, denoted as \textsf{CollaborER-S}.
For fair comparisons with unsupervised methods, we report the results of \textsf{CollaborER} on the whole datasets, denoted as \textsf{CollaborER-U}.

\vspace{0.05in}
\noindent
\textbf{Implementation details.}
We implemented CollaborER\footnote{The source code of CollaborER is available at \url{https://github.com/ZJU-DAILY/CollaborER}} in PyTorch~\cite{pytorch}, the Transformers library~\cite{huggingface}, and the Sentence-Transformers library~\cite{sentenceBERT19}. In automatic label generation (ALG), we use \emph{stsb-roberta-base}\footnote{\url{https://github.com/UKPLab/sentence-transformers}} as the pre-trained LM to get the embedding for every tuple. We set $\theta=0.03$ in the process of reliable positive label generation (RPLG) and $\epsilon=10$\footnote{To avoid false negative labels, we dismiss the top-2 neighbors into consideration.} in the process of similarity-based negative label generation (SNLG). In collaborative ER training (CERT), the dimension of the graph feature in the process of multi-relational graph feature learning (MRGFL) is 128.
Besides, in both the training and test process of \textsf{CollaborER}, we apply the half-precision floating-point (fp16) optimization to save the GPU memory usage and the running time.
In all experiments, the max sequence length is set to 256;
the learning rate is set to 2e-5; the batch size for the AG benchmark is set to 64 while that for the other benchmarks is set to 32.
%We set the batch size to 64 for the AG benchmark and 32 for the other benchmarks.
The training process runs a fixed number of epochs (1, 2, 3, 6, or 30 depending on the dataset size), and returns the checkpoint at the last epoch.
We set $\lambda=0.5$ and $\mu=0.2$ in the proposed \emph{CollaborER training loss}.
All the experiments were conducted on a personal computer with an Intel Core i9-10900K CPU, an NVIDIA GeForce RTX3090 GPU, and 128GB memory. The programs were all implemented in Python.

\begin{figure*}[t]
\centering
\hspace*{5mm}
\includegraphics[width=0.95\textwidth]{experiment_fig/ablation-icon.eps}\\
\vspace*{-1mm}
\subfigure[AG]{
 \includegraphics[width=1.7in]{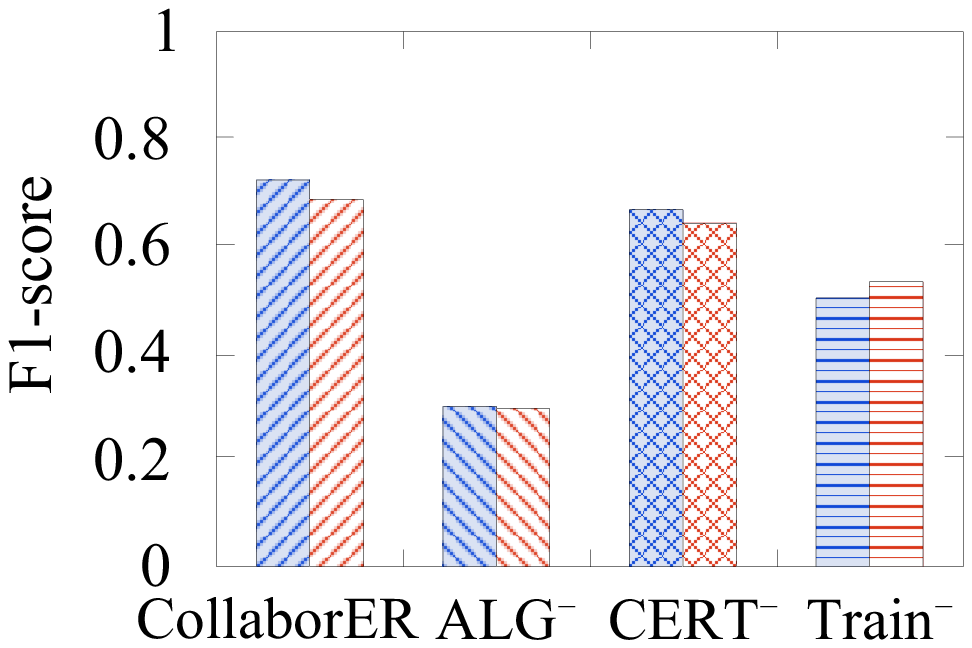}
}\hspace*{-1mm}
\subfigure[BR]{
 \includegraphics[width=1.7in]{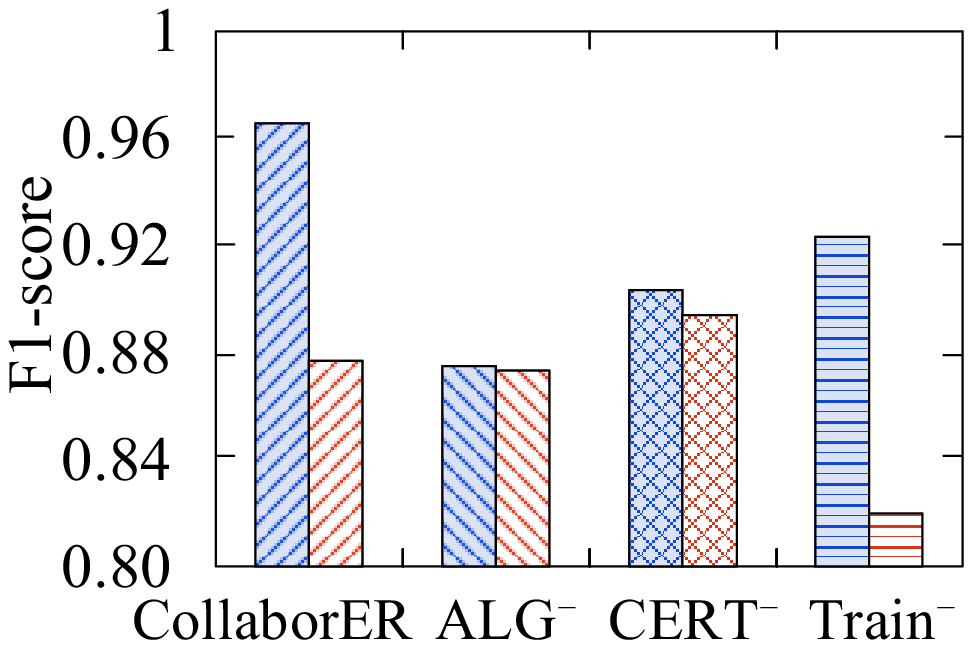}
}\hspace*{-1mm}
\subfigure[DA-clean]{
 \includegraphics[width=1.7in]{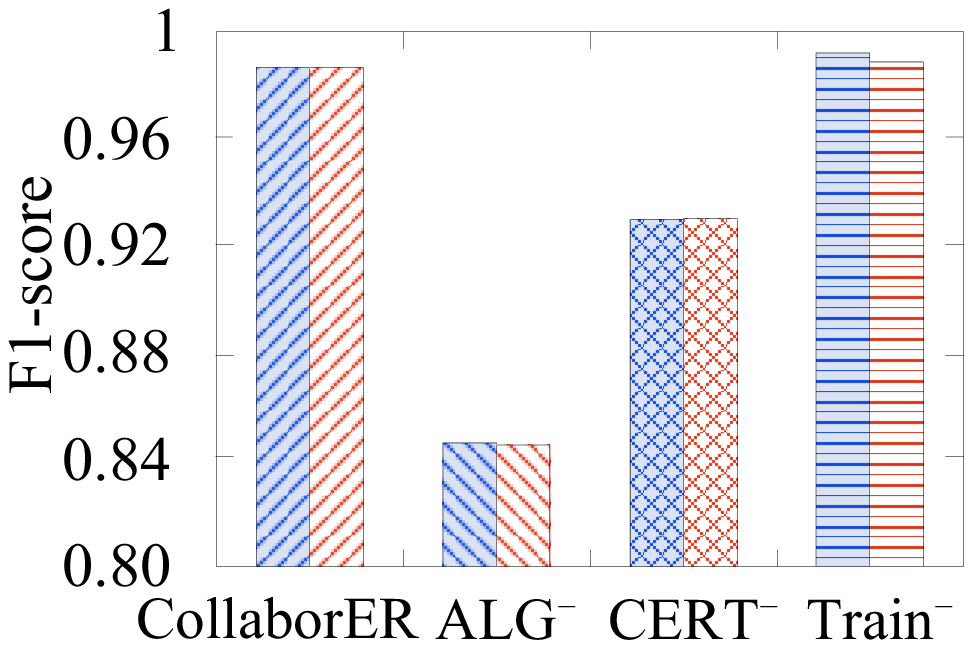}
}\hspace*{-1mm}
\subfigure[FZ]{
 \includegraphics[width=1.7in]{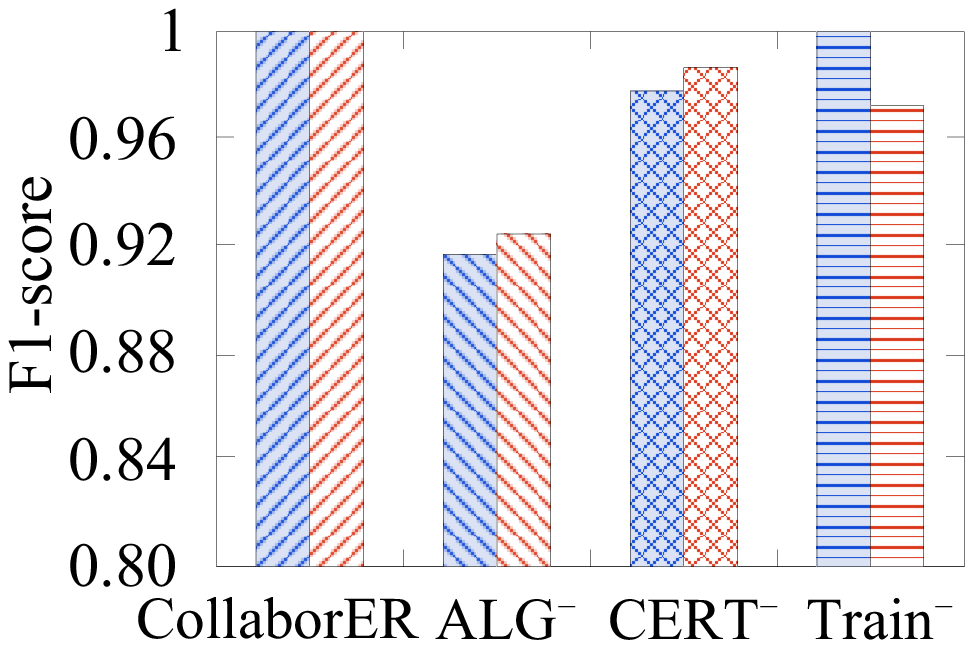}
}\hspace*{-1mm}\\
\subfigure[IA-clean]{
 \includegraphics[width=1.7in]{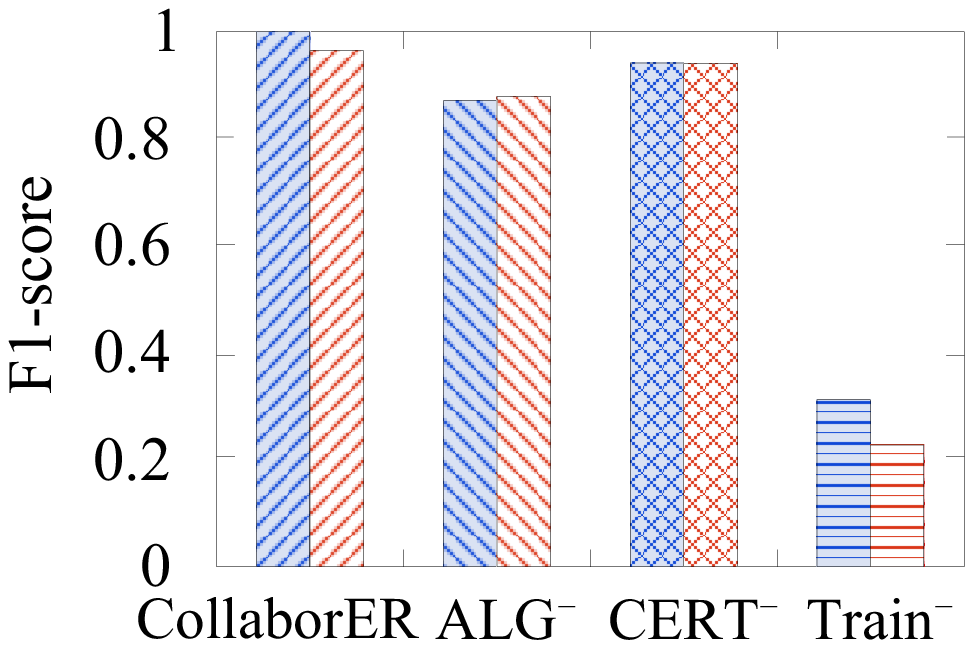}
}\hspace*{-1mm}
\subfigure[DA-dirty]{
 \includegraphics[width=1.7in]{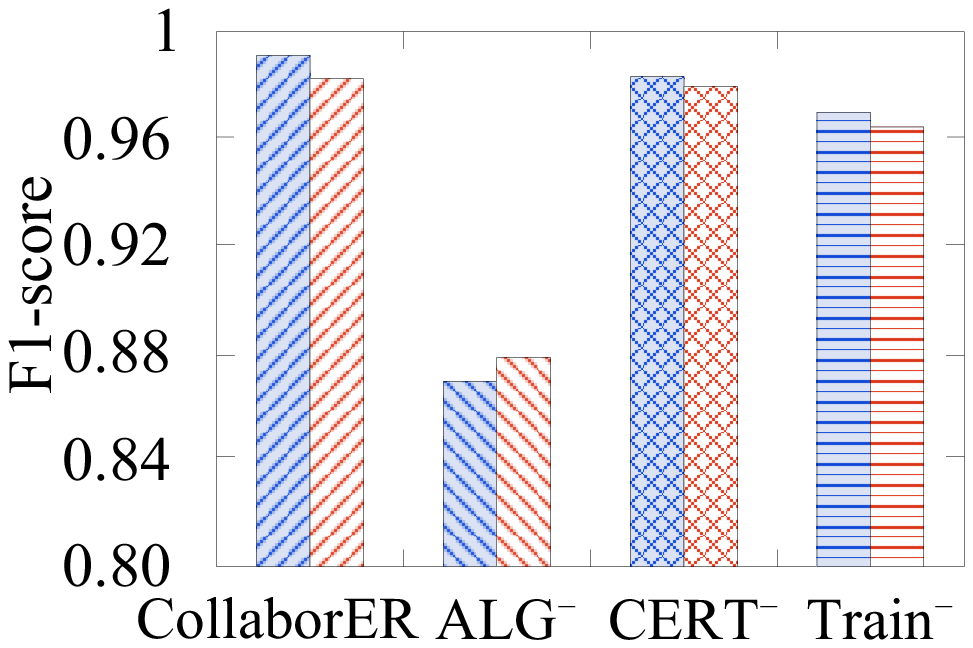}
}\hspace*{-1mm}
\subfigure[IA-dirty]{
 \includegraphics[width=1.7in]{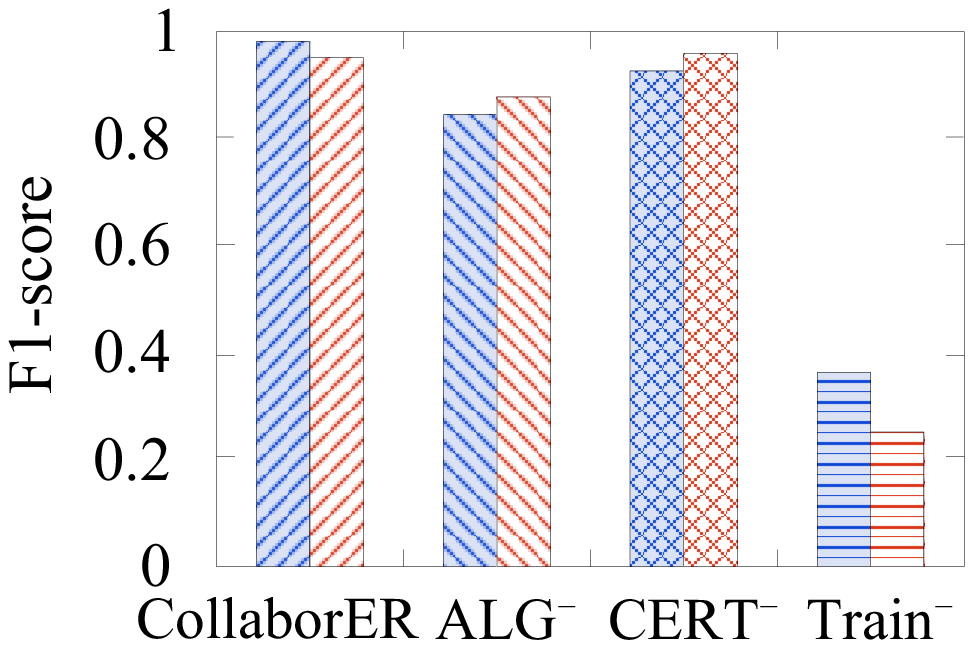}
}\hspace*{-1mm}
\subfigure[AB]{
 \includegraphics[width=1.7in]{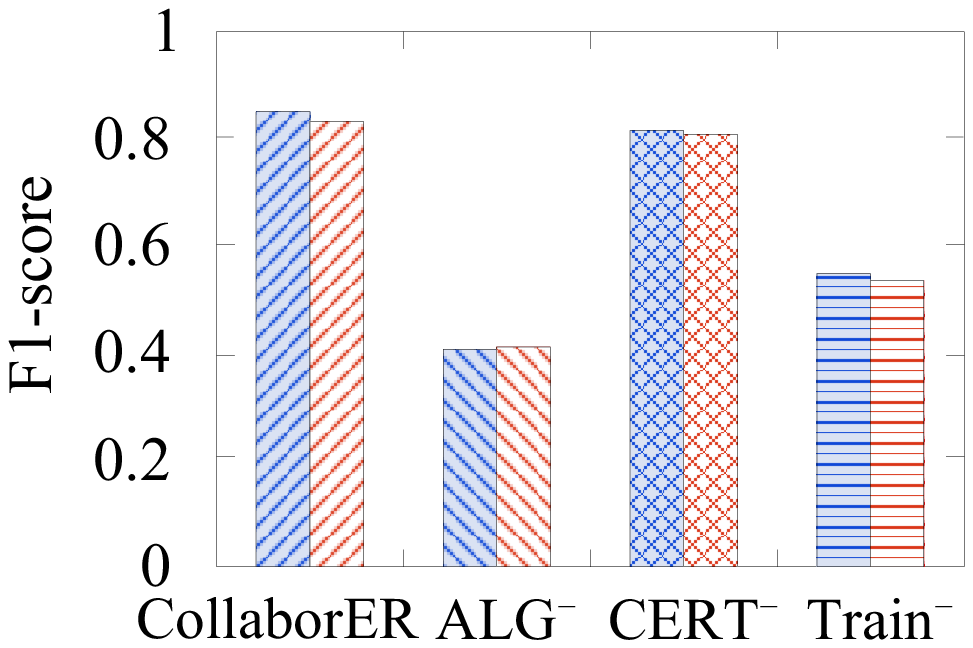}
}
\vspace*{-2mm}
\caption{Ablation study of \textsf{CollaborER}. Blue bars (corresponding to test) represent the results on the test sets, and red bars (corresponding to all) denote the results on the whole datasets.}
\label{fig:ablation}
% \vspace*{-3mm}
\end{figure*}

\subsection{Overall Performance}
\label{exp:overall}

Table~\ref{tb:overall_ER_results} summarizes the overall ER performance of \textsf{CollaborER} and its competitors.

\vspace{0.05in}
\noindent
\textbf{\textsf{CollaborER} vs. unsupervised methods.}
It is observed that \textsf{CollaborER} significantly outperforms all the unsupervised competitors.
Particularly, \textsf{CollaborER} brings about 1\%-85\% absolute improvement over the best baseline (i.e., EMBDI).
The results also demonstrate that \textsf{CollaborER} is more robust against data noise than ZeroER.
On the dirty datasets, the performance degradation of \textsf{CollaborER} is only 0.67\% on average.
Nevertheless, the performance of ZeroER decreases by 33\%.
The reason is that unsupervised methods can easily be fooled without the guidance of any supervision signal, as discussed in Section~\ref{sec:introduction}.
On the contrary, \textsf{CollaborER} generates reliable labels via the proposed ALG strategy as the supervision signals.
The reliability analysis of ER labels generated by ALG can be found in Section~\ref{exp:ALG_analysis}.
Besides, the collaborative ER training process (i.e., CERT), which absorbs both graph features and sentence features of tuples, has the fault-tolerance capability for dealing with noisy tuples.
The outstanding ER performance and the robust property make \textsf{CollaborER} more attractive in practical ER scenarios.

\vspace{0.05in}
\noindent
\textbf{\textsf{CollaborER} vs. supervised methods.}
As we can see, the performance of \textsf{CollaborER} is comparable with or even superior to the SOTA supervised ER approaches.
Concretely, \textsf{CollaborER} outperforms even the best supervised competitor (i.e., DITTO) by 1.54\% on average over 5 datasets.
Although the performance of \textsf{CollaborER} in the other three datasets is inferior to that of DITTO, the difference in their respective F1-scores does not exceed 4\%.
This is really impressive since \textsf{CollaborER} requires \emph{zero} human involvement in annotating labels for ER.
In contrast, DITTO requires a sufficient amount of labels that are expensive to obtain and often times infeasible.
To further compare the performance difference between \textsf{CollaborER} and DITTO under a fair comparison, we evaluate the performance of DITTO when using the pseudo labels generated by ALG, denoted as DITTO-S. As can be observed, the performance of DITTO-S is inferior to that of \textsf{CollaborER}. The reason is that, DITTO-S treats entities as sentences, and hence, it does not consider the rich semantic features of entities, as mentioned in Section~\ref{sec:introduction}. Nonetheless, \textsf{CollaborER} has the capability to capture rich semantics of entities by discovering both sentence features and graph features of entities collaboratively.

In addition, both \textsf{CollaborER} and ERGAN generate pseudo labels for the purpose of improving their ER performance. However, we can observe from the results that \textsf{CollaborER} achieves superior performance than ERGAN.
%In addition, one may be curious about the reason why \textsf{CollaborER} achieves superior performance than ERGAN, since they both generate pseudo labels for the purpose of improving their ER performance.
Specifically, \textsf{CollaborER} brings about 23\% improvement on average on the F1-score, compared to the ER results produced by ERGAN.
The inferior performance of ERGAN is attributed to the inherent GAN. To be more specific, it is common that the training process of GAN is unstable~\cite{MetzPPS17}, incurring the poor quality of the generated pseudo labels and unsatisfied training performance. By analyzing the evaluated datasets, ERGAN generates positive and negative labels with an average accuracy of 88\% and 89\%, respectively. However, \textsf{CollaborER} produces positive and negative labels with an average accuracy of 99\% and 97\% respectively, as verified in Section~\ref{exp:ALG_analysis}. Compared to ERGAN, \textsf{CollaborER} gains up to 11\% improvement on the quality of the generated labels.

\subsection{Ablation Study}\label{sec:ablation}

Next, we analyse the effectiveness of each proposed phase of \textsf{CollaborER} (i.e., ALG and CERT) by comparing \textsf{CollaborER} with its variants without the key optimization(s) in each phase. The results are shown in Figure~\ref{fig:ablation},
where the labels listed along the abscissa have the following meanings: (i) ``CollaborER'' represents its performance when all optimizations are used; (ii) ``ALG$^-$'' means the performance of \textsf{CollaborER} without (w/o) ALG; (iii) ``CERT$^-$'' denotes the performance of \textsf{CollaborER} w/o CERT; and (iv) ``Train$^-$'' represents the performance of \textsf{CollaborER} w/o training.
% \baihua{Congcong, can we change the labels to include superscript $^{-}$ (e.g., ALG $\rightarrow$ ALG$^-$) to make them more straightforward? For 'ALL', I will suggest to just use \textsf{CollaborER}.}

% \baihua{Need to explain how to interpret the results shown in Figure~\ref{fig:ablation}. Does NONE mean \textsf{CollaborER}? Legends and the labels listed along x-dimension are inconsistent. }

\vspace{0.05in}
\noindent
\textbf{\textsf{CollaborER} vs. \textsf{CollaborER} w/o ALG.}
ALG contains two components, i.e., RPLG and SNLG.
Since \textsf{CollaborER} cannot work without RPLG,
we focus on investigating the effectiveness of SNLG by replacing it with a random negative label generation method.
It is observed that the F1-score drops 17.87\% on average.
This confirms that generating ``challenging'' negative labels based on semantic similarity greatly helps to train effective ER models.
We also observe that the SNLG brings no more than 8.33\% improvement on FZ dataset.
This is attributed to the nature of this dataset, as it is relatively easier for \textsf{CollaborER} and all the competitors to achieve the perfect performance, i.e., 100\% F1-score, in this dataset.
% For instance, on the DA-clean dataset, it dramatically drops by \textcolor{red}{$\sim xx\%$} on F1-score.

\vspace{0.05in}
\noindent
\textbf{\textsf{CollaborER} vs. \textsf{CollaborER} w/o CERT.}
The difference between the proposed CERT and other existing ER models that also fine-tune pre-trained LMs lies in whether there is the intervention of graph features (w.r.t MRGFL) to assist the fine-tuning process.
By removing MRGFL in CERT, the F1-score of \textsf{CollaborER} drops 3\% on average over the eight experimental datasets.
Particularly, the drop of the F1-score is up to 5\% on the DA-clean dataset.
This shows that learning tuples' graph features is indispensable for promoting ER performance.

\vspace{0.05in}
\noindent
\textbf{\textsf{CollaborER} vs. \textsf{CollaborER} w/o Train.}
We also explore the performance of \textsf{CollaborER} without any training process.
In this case, \textsf{CollaborER} performs ER purely based on RPLG, which automatically discovers the matched tuples based on the semantic similarity.
The results indicate that RPLG can find a large quantity of reliable matched tuples.
It is worth noting that RPLG alone can achieve considerable results, e.g., $\sim$~99\% F1-score and 100\% F1-score on DA-clean dataset and FZ dataset, respectively.
This is because, matched tuples are mutually most similar with each other in those datasets.
Since RPLG is general enough to perform ER in various datasets, it is possible to be widely used in practical ER applications without any time-consuming training process.

% \begin{table*}[t]
% \caption{Ablation Study}
% \setlength{\tabcolsep}{2.15mm}{
% \begin{tabular}{c|ccccc|cc|c}
% \toprule
% \multirow{2}{*}{Methods} & \multicolumn{5}{c|}{Structured}  & \multicolumn{2}{c|}{Dirty} & Textual \\
%  & AG & BR & DA-clean & FZ & IA-clean & DA-dirty & IA-dirty   & AB  \\ \hline
% CollaborER   & & & \textbf{} & &   &  & & \\
% CollaborER w/o SNLG  & & &   & &   &  & & \\
% CollaborER w/o CERT  & & & \textbf{} & &   &  & & \\
% CollaborER w/o Training  & & & \textbf{} & &   &  & & \\ \bottomrule
% \end{tabular}}
% \end{table*}

\subsection{Further Experiments}
We further justify the effectiveness of the proposed \textsf{CollaborER} by conducting the following two sets of experiments.

\subsubsection{Graph Scale Analysis}\label{exp:HGC_analysis}

\begin{figure}[t]
    \centering
    \hspace{2mm}
    \includegraphics[width=0.5\textwidth]{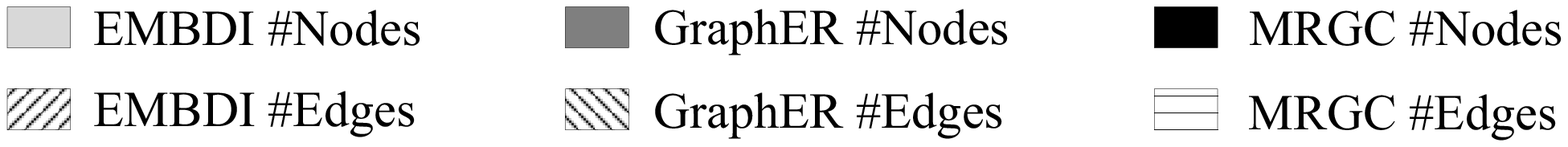}\\
    \vspace{1mm}
    \centering
    \hspace{-3.4mm}
    \includegraphics[width=3.6in]{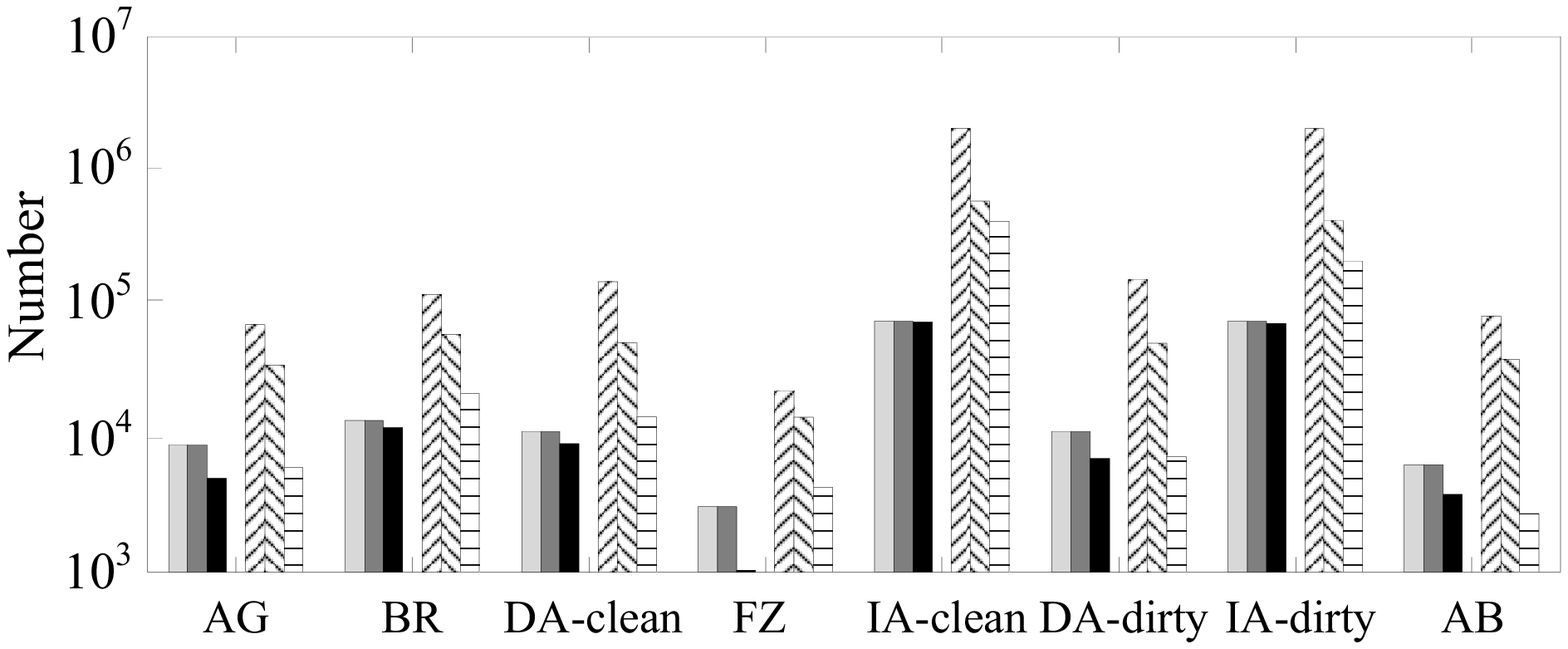}
    \vspace{-7mm}
    \caption{Graph scale analysis}
    \label{fig:graph-scale-analysis}
    \vspace{-3mm}
\end{figure}

The first set of experiments is to compare the scale of the graphs generated by the proposed MRGC and other graph construction methods in the existing ER approaches, i.e., EMBDI~\cite{CreatingEmDI20} and GraphER~\cite{GraphER20}.
Figure~\ref{fig:graph-scale-analysis} depicts the total number of nodes (denoted as \#Nodes) and that of edges (denoted as \#Edges) of graphs with regard to each dataset.
It is observed that MRGC generates much smaller graphs, compared against other graph generation methods.
The reduced size of graphs greatly saves the memory for storing and training graphs, and reduces the training cost. 

\begin{table}[!]
\caption{The reliability analysis of ALG}
\vspace{-2mm}
\setlength{\tabcolsep}{2.8mm}{
\begin{tabular}{c|ccc|ccc}
\toprule
\multirow{2}{*}{\textbf{Datasets}} & \multicolumn{3}{c|}{\textbf{RPLG}} & \multicolumn{3}{c}{\textbf{SNLG}} \\ \cline{2-7}
  & \textbf{TP}  & \textbf{FN}   & \textbf{TPR}  & \textbf{TN}  & \textbf{FP} & \textbf{TNR} \\ \hline
AG & 332 & 0 & 1   & 7136 & 115 & 0.98  \\
BR & 34  & 0 & 1   & 17417   & 1074   & 0.94  \\
DA-clean & 2136 & 0 & 1   & 34119   & 3  & 0.99  \\
FZ & 102 & 0 & 1   & 1788 & 11 & 0.99  \\
IA-clean & 4   & 0 & 1   & 8399 & 521 & 0.94  \\
DA-dirty & 1941 & 0 & 1   & 30899   & 7  & 0.99  \\
IA-dirty & 2   & 0 & 1   & 7872 & 490 & 0.94  \\
AB & 247 & 2 & 0.99 & 4093 & 11 & 0.99  \\ \bottomrule
\end{tabular}}\label{tb:ALG_quality}
\vspace{-3mm}
\end{table}

\begin{figure*}[t]
\centering
% \vspace{-2mm}
\hspace*{3mm}
\includegraphics[width=0.85\textwidth]{experiment_fig/parameter-icon.eps}\\
\vspace*{-2mm}
\subfigure[AG]{
 \includegraphics[width=1.8in]{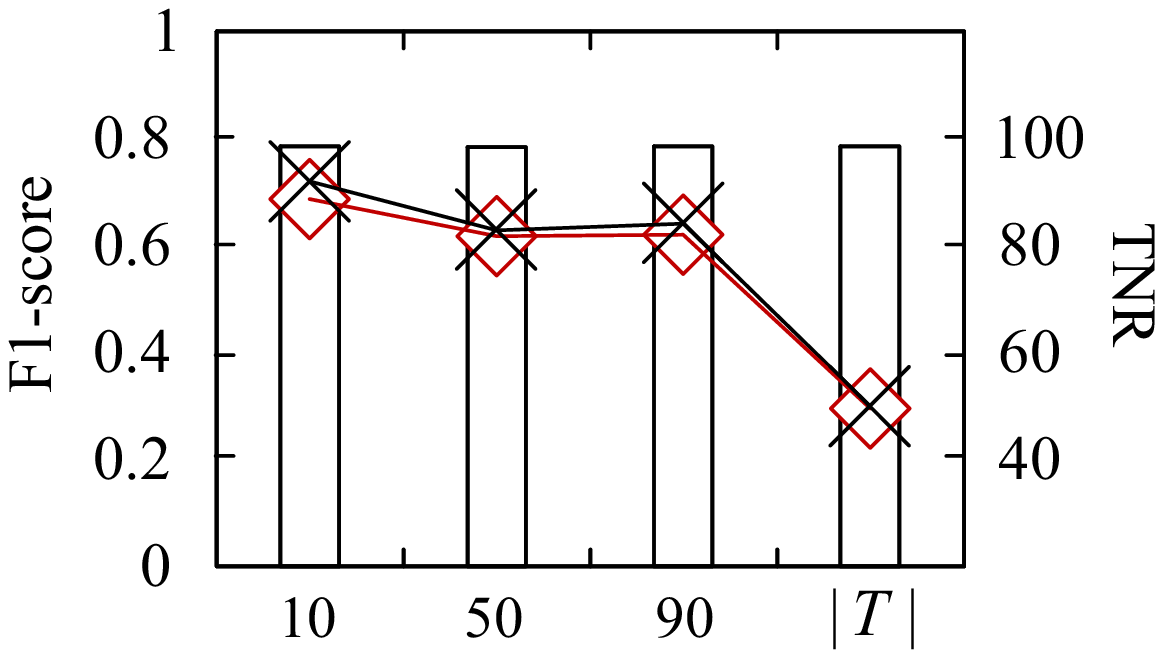}
}\hspace*{-3mm}
\subfigure[BR]{
 \includegraphics[width=1.8in]{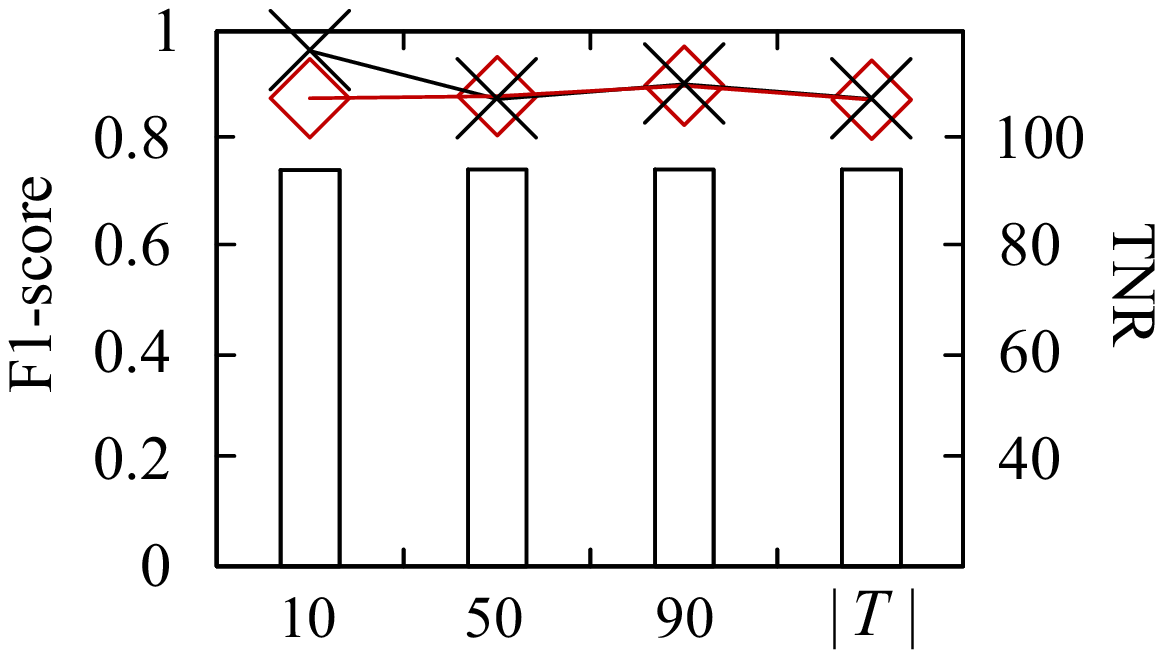}
}\hspace*{-3mm}
\subfigure[DA-clean]{
 \includegraphics[width=1.8in]{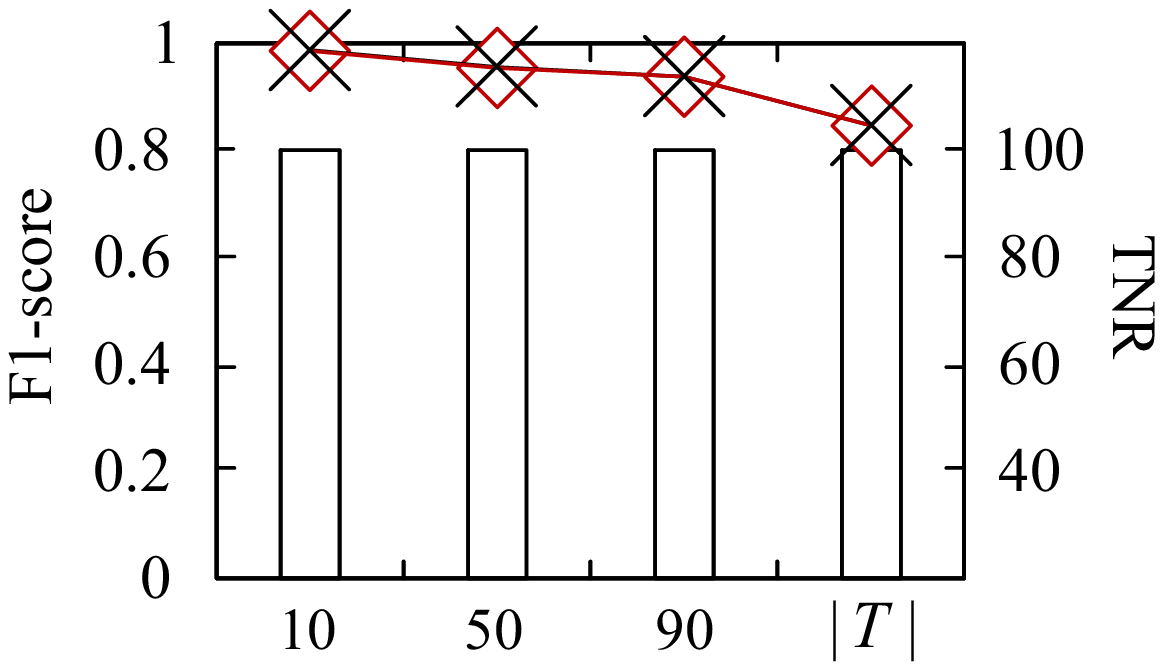}
}\hspace*{-3mm}
\subfigure[FZ]{
 \includegraphics[width=1.8in]{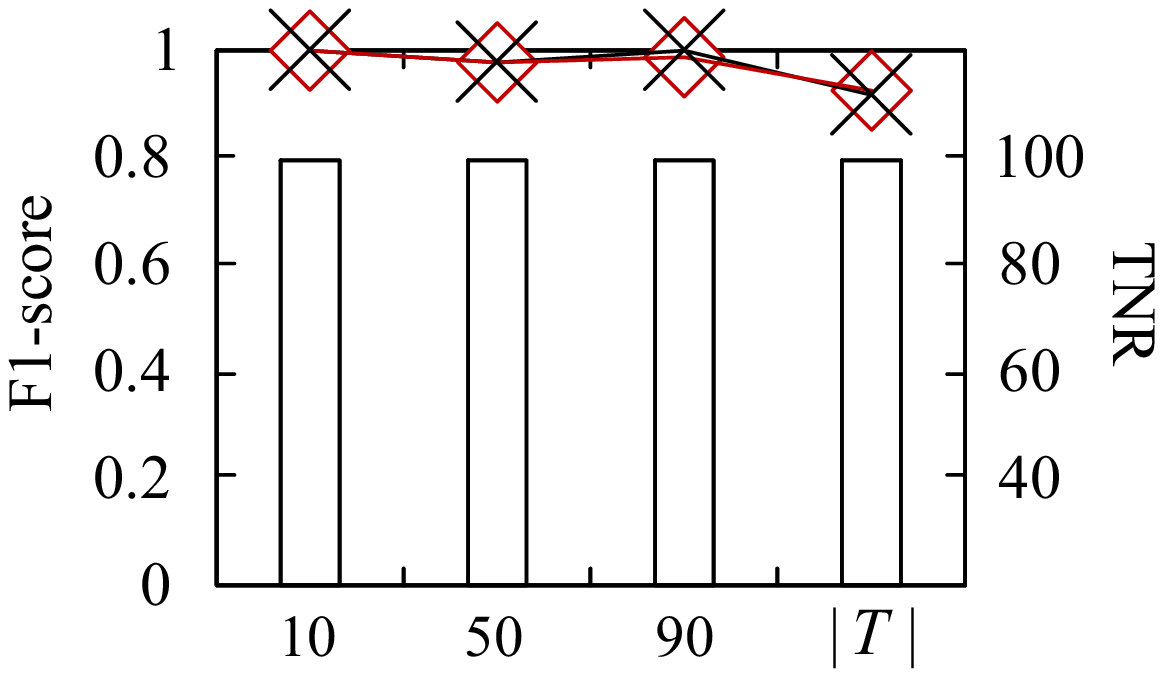}
}\vspace*{-1mm}\\
\subfigure[IA-clean]{
 \includegraphics[width=1.8in]{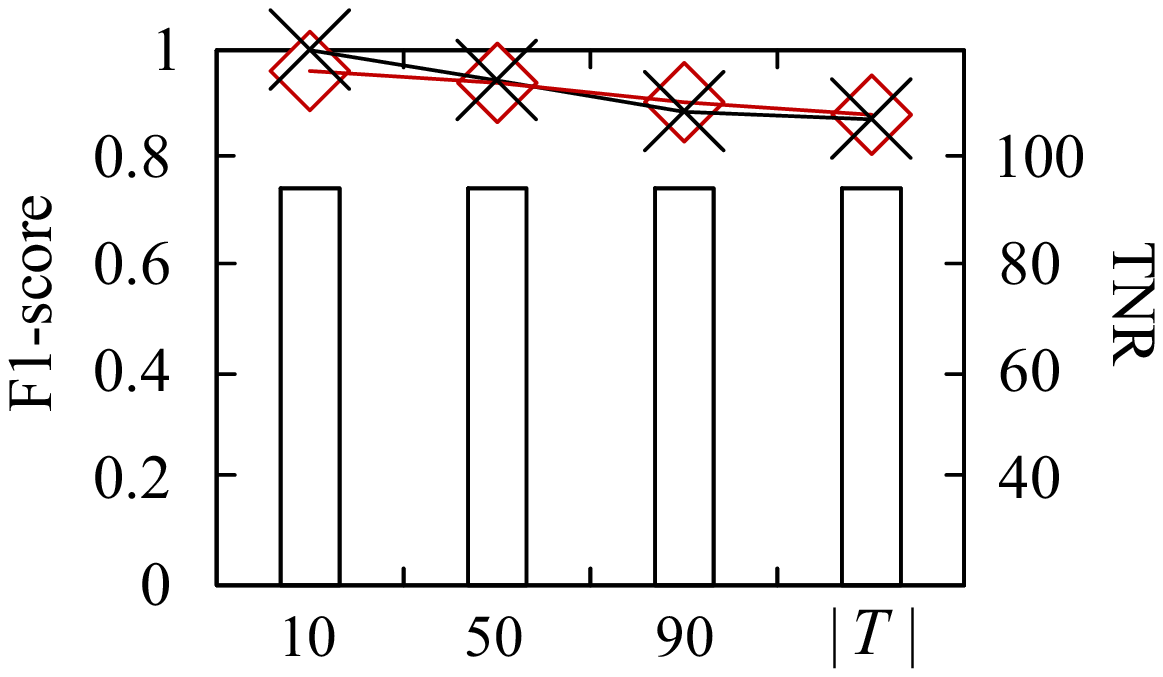}
}\hspace*{-3mm}
\subfigure[DA-dirty]{
 \includegraphics[width=1.8in]{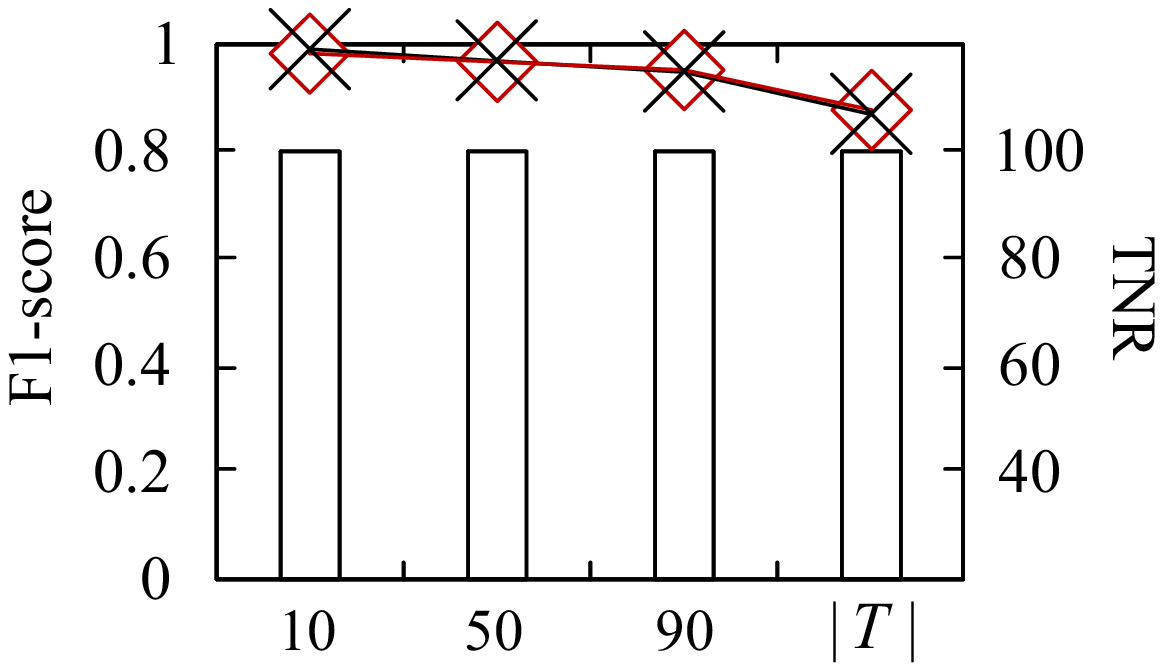}
}\hspace*{-3mm}
\subfigure[IA-dirty]{
 \includegraphics[width=1.8in]{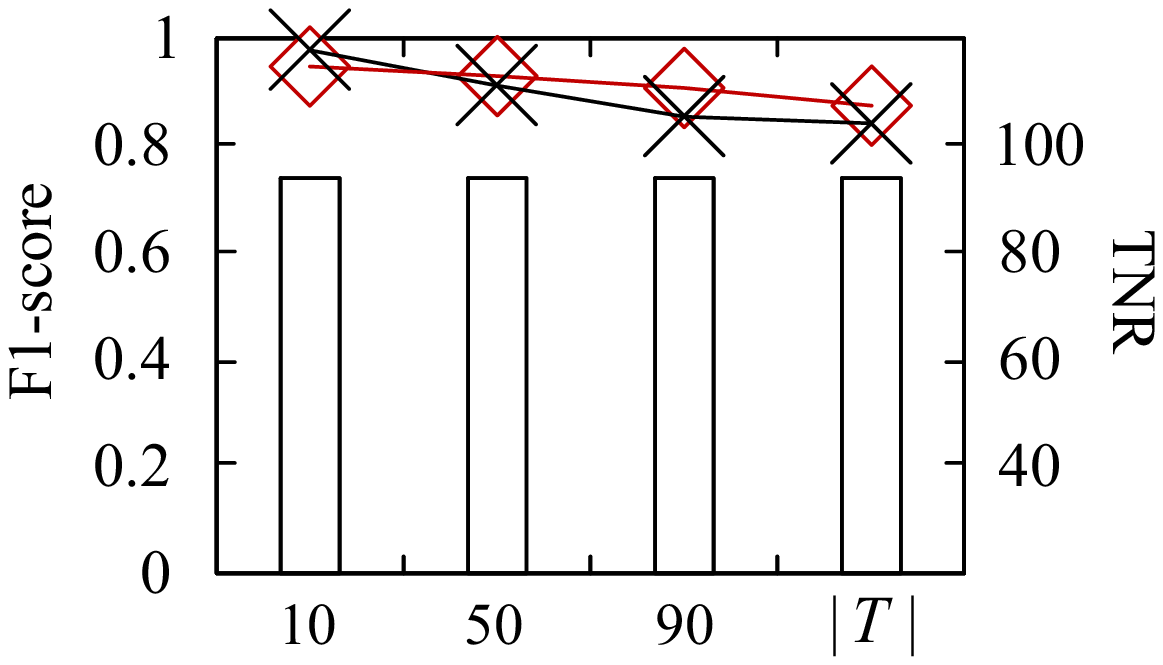}
}\hspace*{-3mm}
\subfigure[AB]{
 \includegraphics[width=1.8in]{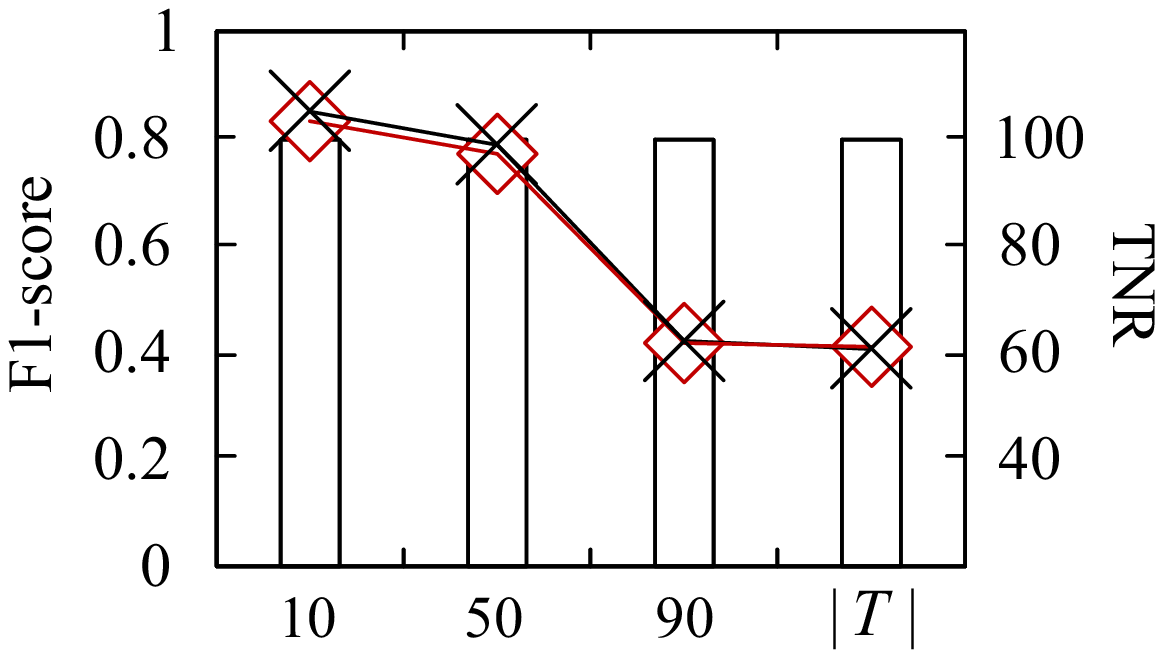}
}
\vspace*{-3mm}
\caption{Effect of $\epsilon$-nearest neighbors for SNLG}
\label{fig:effect-of-nearest-for-snlg}
\vspace*{-5mm}
\end{figure*}

\subsubsection{ALG Analysis}
\label{exp:ALG_analysis}

The second set of experiments is to verify the performance of ALG.
To better study the quality of labels, we utilize six metrics:
(i) \emph{true-positive} (TP), which represents the number of truly labeled matched tuples;
(ii) \emph{true-negative} (TN), which denotes the number of truly labeled mismatched tuples;
(iii) \emph{false-negative} (FN), which represents the number of matched tuples that are labeled as mismatched;
(iv) \emph{false-positive} (FP), which denotes the number of mismatched tuples that are labeled as matched;
(v) \emph{true-positive rate} (TPR) represents the proportion of matched tuples that are correctly labeled, denoted as $\frac{TP}{TP+FN}$; and
(vi) \emph{true-negative rate} (TNR) represents the proportion of mismatched tuples that are correctly labeled, denoted as $\frac{TN}{TN+FP}$.

\vspace{0.05in}
\noindent
\textbf{Analysis of label generating quality.}
We first evaluate the quality of the labels generated by ALG, including the positive labels generated by PRLG and the negative labels produced by SNLG.
The results are reported in Table~\ref{tb:ALG_quality}.
As expected, both PRLG and SNLG are able to achieve the outstanding performance when generating labels. Specifically, \textsf{CollaborER} produces positive and negative labels with an average accuracy of 99\% and 97\%, respectively.
It confirms the effectiveness of our proposed ALG.
The positive labels with high reliability allow the subsequent CERT model to be well-trained; while
the generated negative labels enable CERT to identify ``challenging'' tuple pairs.

\vspace{0.05in}
\noindent
\textbf{Effect of $\epsilon$-nearest neighbors for SNLG.}
We then study the performance of SNLG by varying $\epsilon$ among \{10, 50, 90, $|T|$\}.
Note that, when $\epsilon = |T|$, SNLG generates negative labels by searching for possible tuples in the entire dataset.
In this case, SNLG behaves like random sampling.
%the performance of SNLG is equal to that of random sampling.
Figure~\ref{fig:effect-of-nearest-for-snlg} plots the corresponding results.
We observe that F1-score of \textsf{CollaborER} drops as $\epsilon$ grows.
This is because, the larger the $\epsilon$, the more likely the tuple pairs that are not similar to each other will be included in the set of negative labels.
The dissimilar tuples contribute little to the training of an effective ER model, as discussed in Section~\ref{sec:framework}.
Besides, as expected, the quality of negative labels is still stable when $\epsilon$ changes, which could be observed from TNR (true-negative rate) values.
This further demonstrates the effectiveness of the proposed SNLG.

\section{Demonstration}
\label{sec:demo}

\begin{figure}[t]
    \centering
    \includegraphics[width=3.5in]{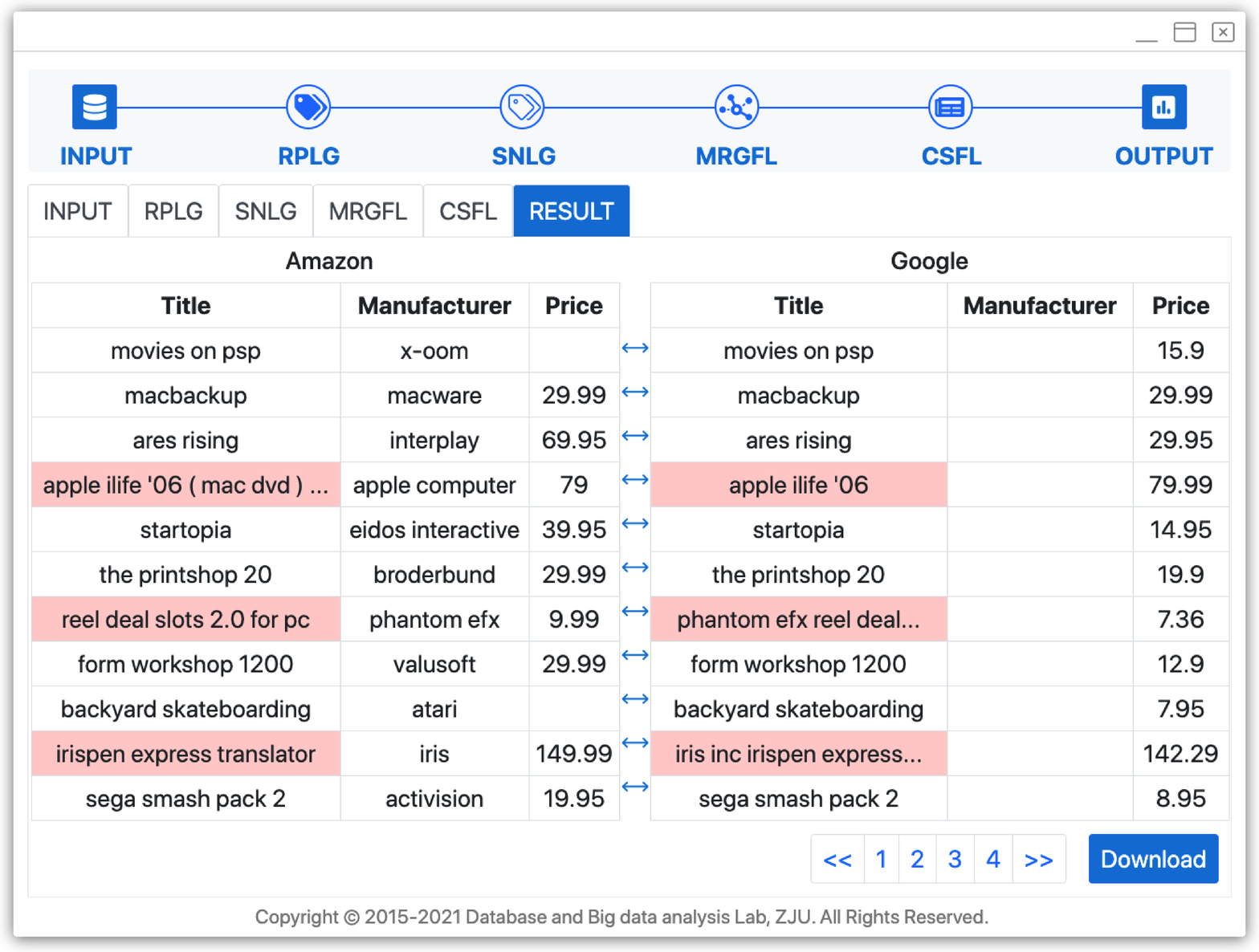}
    \vspace{-4mm}
    \caption{Demonstration. The detected anomalous values are highlighted in red cells.}
\label{fig:demonstration}
\vspace{-4mm}
\end{figure}

Based on the proposed \textsf{CollaborER}, we further develop a cross-platform prototype system \textsf{CollaborAD} for anomaly detection. The basic interface is shown in Figure~\ref{fig:demonstration}.
The upper navigation bar illustrates the main components of the demo system, including the input component, the ER component, and the output component.
We demonstrate \textsf{CollaborAD} using a real-world ER benchmark, i.e., Amazon-Google.
At the input, a user receives two datasets from Amazon-Google. 
%\baihua{receive the datasets from where? Why a dataset must contain anomaly? Ideally, if there is no anomaly in both datasets, the system does not highlight any mistake.}
After the system pre-processes the datasets, it proceeds to perform ER task.
During ER, the system first generates positive labels and negative labels via RPLG and SNLG, respectively.
Then, it constructs graphs for the inputs and learns the graph features of tuples according to the graph structures via MRGFL. Next, it employs the well-trained graph features of tuples to assist the CSFL model in discovering sufficient sentence features of tuples, and produces the final ER results.
As depicted in Figure~\ref{fig:demonstration}, 
%the current progress is in the output part. The
the ER results and the identified anomalies are reported by the output component.
It displays a set of matched tuples connected by double-head arrows. 
The anomalous values of those tuples are highlighted in red cells.
In the event that there is no anomaly in any of the datasets, the system does not highlight anything.

\section{Related Work}\label{sec:relatedwork}

Entity Resolution (ER) is one of the fundamental and significant tasks in data curation.
Early studies exploit rules~\cite{DNF95, ArasuRS09, FanJLM09, MEMPQS017, SinghMEMPQST17} or crowdsourcing~\cite{GokhaleDDNRSZ14, MarcusWKMM11, crowdER12} for ER tasks.
Rule-based solutions require human-provided declarative matching rules~\cite{DNF95} or program-synthesized matching rules~\cite{SinghMEMPQST17} to find matching pairs.
Crowdsourcing-based solutions employ crowds to manually identify whether two tuples refer to the same real-world entity.
Such solutions highly rely on human guidance, and have limitations in handling heterogeneous data.
Recently, machine learning (ML) techniques have been widely used for ER and have achieved promising performance~\cite{DIsurvey18}.
According to whether supervision signals are incorporated, existing ML-based solutions can be clustered into two categories, namely, \emph{supervised ER} and \emph{unsupervised ER}.

Supervised ER approaches~\cite{DeepER18, KopckeTR10, DeepMatcher18, Ditto20, GraphER20, AutoEM19, ERIJCAI19, ActiveTransferER19, bertER2021, MCA20, BrunnerS20, CreatingEmDI20} can provide the state-of-the-art performance for ER, but require a substantial number of labels in the form of matches and mismatches, to support the learning of a reliable ER model.
In general, the methods first learn the features of tuples via ML models and then feed the well-trained features into a binary classifier for identifying matched tuples.

A majority of supervised ER methods employ sentence-based ML model to learn the \emph{sentence features} of tuples.
DeepER~\cite{DeepER18} and DeepMatcher~\cite{DeepMatcher18} utilize vanilla RNNs.
MCA~\cite{MCA20} proposes a multi-context attention mechanism to enrich the sentence features of tuples.
Furthermore, current studies~\cite{Ditto20, BrunnerS20, bertER2021} indicate that applying pre-trained LMs to ER tasks achieves outstanding performance.
DITTO~\cite{Ditto20} obtains the best performance among all the existing supervised ER works.
It fine-tunes the pre-trained LMs with the help of a series of newly proposed data augmentation techniques.
Several supervised ER methods transform a collection of tuples with the relational format to graph structures, and learn the \emph{graph features} of tuples based on the constructed graphs~\cite{GraphER20, CreatingEmDI20}.

However, both sentence-based methods and graph-based methods are far from enough to capture sufficient features of tuples, as mentioned in Section~\ref{sec:introduction}.
Our proposed \textsf{CollaborER} is introduced to enrich the features of tuples by learning both sentence features and graph features collaboratively.
Besides, we have compared \textsf{CollaborER} with \emph{four} state-of-the-art supervised ER solutions, and have verified that \textsf{CollaborER}, with zero labor-intensive labeling process, achieves comparable or even superior results, as compared with supervised approaches.

Unsupervised ER approaches~\cite{ZeroER20, CreatingEmDI20, zhang2020unsupervised} 
are designed to perform ER without labeling.
ZeroER~\cite{ZeroER20} learns the match and mismatch distributions based on Gaussian Mixture Models.
EMBDI~\cite{CreatingEmDI20} performs ER by learning a compact graph-based representation for each tuple.
ITER+CliqueRank~\cite{zhang2020unsupervised} first constructs a bipartite graph to model the relationship between tuple pairs, and then develops an iterative-based ranking algorithm to estimate the similarity of tuple pairs.
Despite the benefit of zero label requirement, unsupervised approaches are highly error-sensitive and may suffer from poor ER results when errors are contained in datasets.
Considering that real-world datasets are often dirty, it is impractical to use the existing unsupervised ER methods in practice.

On the contrary, the proposed \textsf{CollaborER} performs ER in a self-supervised manner, which has the capability to perform ER in a fault-tolerant manner, as verified in the experiments reported in Section~\ref{exp:overall}.
In addition, we have compared \textsf{CollaborER} with two state-of-the-art unsupervised methods, including ZeroER and EMBDI.
%We do not compare \textsf{CollaborER} with
Note that we exclude ITER+CliqueRank from experiments since its performance is inferior to the two unsupervised methods that are selected as competitors in our study.
\section{Conclusions}\label{sec:conclusions}

In this paper, we propose \textsf{CollaborER}, a self-supervised entity resolution framework, to perform the ER task with \emph{zero} labor-intensive manual labeling.
\textsf{CollaborER} conducts ER tasks by a pipe-lined modular architecture consisting of two phases, i.e., \emph{automatic label generation} (ALG) and \emph{collaborative ER training} (CERT).
First, ALG is developed to automatically generate both \emph{reliable positive labels} (w.r.t. RPLG) and \emph{semantic-based negative labels} (w.r.t. SNLG).
ALG is essential for the subsequent CERT phase since it provides high-quality labels that are the backbone of training effective ER models.
Second, the framework proceeds to the CERT phase, where tuples' sentence features and graph features are learned and employed collaboratively to produce the final ER results.
In this phase, we first propose a \emph{multi-relational graph construction} (MRGC) method to construct graphs for each relational dataset, and then exploit GNN to learn the graph features of tuples.
Thereafter, the well-trained graph features are fed into \emph{a collaborative sentence feature learning} (CSFL) model to discover sufficient sentence features of tuples.
Finally, CSFL predicts the matched tuple pairs and unmatched ones according to the learned features.

Currently, \textsf{CollaborER} treats the graph feature learning process (i.e., CSFL) as a black box, and uses graph features of tuples generated by AttrGNN~\cite{AttrGNN20} in the current implementation. In the near future, we plan to design a graph feature engineering solution for ER. 
%is an interesting future work.

\balance

\bibliographystyle{abbrv}
\bibliography{refer}

\end{document}